\documentclass[a4paper,fleqn,usenatbib]{mnras}

\usepackage{newtxtext,newtxmath}

\usepackage[T1]{fontenc}
\usepackage{ae,aecompl}

\usepackage{graphicx}	
\usepackage{amsmath}	
\usepackage{amssymb}	
\usepackage{pdflscape}  
\usepackage{adjustbox}  
\usepackage{enumerate}  
\usepackage{float}      
\usepackage{caption}
\usepackage{nicefrac}
\usepackage{siunitx}
\usepackage{enumerate}

\newcommand\Tstrut{\rule{0pt}{1.5ex}}       
\newcommand\Bstrut{\rule[-1.5ex]{0pt}{0pt}}

\title[Extreme variability of SDSS J2232$-$0806]{The `Big Dipper': The nature of the extreme variability of the AGN SDSS\,J2232$-$0806}

\author[D. Kynoch et al.]{Daniel Kynoch$^{1}$\thanks{E-mail: daniel.kynoch@durham.ac.uk},
Martin J.\ Ward$^{1}$,
Andy Lawrence$^{2}$,
Alastair G.\ Bruce$^{2}$,
\newauthor
Hermine Landt$^{1}$ and
Chelsea L.\ MacLeod$^{3}$
\\
$^{1}$Centre for Extragalactic Astronomy, Department of Physics, Durham University, South Road, Durham, DH1 3LE, UK\\
$^{2}$Institute for Astronomy, SUPA (Scottish Universities Physics Alliance), University of Edinburgh, Royal Observatory, Blackford Hill, \\ Edinburgh EH9 3HJ, UK \\
$^{3}$Harvard-Smithsonian Center for Astrophysics, 60 Garden Street, Cambridge, MA 02138, USA
}

\date{Accepted XXX. Received YYY; in original form ZZZ}

\pubyear{2019}

\begin{document}
\label{firstpage}
\pagerange{\pageref{firstpage}--\pageref{lastpage}}
\maketitle

\begin{abstract}
SDSS J2232$-$0806 (the `Big Dipper') has been identified as a `slow-blue nuclear hypervariable': a galaxy with no previously known active nucleus, blue colours and large-amplitude brightness evolution occurring on a timescale of years.
Subsequent observations have shown that this source does indeed contain an active galactic nucleus (AGN).
Our optical photometric and spectroscopic monitoring campaign has recorded one major dimming event (and subsequent rise) over a period of around four years; 
there is also evidence of previous events consistent with this in archival data recorded over the last twenty years.
Here we report an analysis of the eleven optical spectra obtained to date and we assemble a multiwavelength data set including infrared, ultraviolet and X-ray observations.
We find that an intrinsic change in the luminosity is the most favoured explanation of the observations, based on a comparison of continuum and line variability and the apparent lagged response of the hot dust.
This source, along with several other recently-discovered `changing-look' objects, demonstrate that AGN can exhibit large-amplitude luminosity changes on timescales much shorter than those predicted by standard thin accretion disc models. 
\end{abstract}




\section{Introduction}
\label{sec:int}
Active galactic nuclei (AGN) are powered by the gravitational energy reprocessed as matter spirals inward and is finally accreted by the central supermassive black hole (BH).  
Two of the defining characteristics of AGN are their very high bolometric luminosities, and in the case of those that are not obscured, by their significant multi-frequency variability on many timescales. 
Numerous variability studies have been conducted, both on large samples of AGN (e.g.\ Stripe 82, \citealt{MacLeod12}, \citealt{Schmidt12}, and \citealt{Zuo12}) and detailed studies of individual cases (e.g.\ NGC\,4593 by  \citealt{McHardy18} and NGC\,5548 by \citealt{Pei17} and references therein).   
In addition to these studies some cases of extreme variability have been identified in the form of the so-called `changing-look' quasars (CLQs: e.g.\ \citealt{MacLeod18}, \citealt{Yang18}, \citealt{Rumbaugh18} and \citealt{LaMassa15}) which are AGN with (dis)appearing broad emission lines as well as strong continuum changes.  
It is very probable that more than one physical mechanism is responsible for the variations seen across all samples.
Changes in the dust extinction in some AGN were proposed in early studies (e.g.\ \citealt{Goodrich95}), but this explanation is not generally preferred in the case of changing-look AGN. 
In recent studies, often the most favoured cause is a change in the emission from the accretion disc or its associated Comptonisation regions (e.g.\ \citealt{Katebi18}, \citealt{Noda18}, \citealt{Stern18}, \citealt{Ross18}, \citealt{Wang18}, \citealt{Sheng17}, \citealt{Gezari17}, \citealt{Parker16}, \citealt{Ruan16}, \citealt{MacLeod16}, \citealt{Runnoe16} and \citealt{LaMassa15}). 
Other, rarer events, such as stellar tidal disruption, supernovae in the nuclear regions, and gravitational microlensing, have also been proposed (e.g.\ \citealt{Lawrence16}, \citealt{Bruce17} and references therein). 
To make further progress it is important to better characterise the properties of variability to help distinguish between the various mechanisms responsible. 

\subsection{The source SDSS\,J2232$-$0806}
SDSS\,J223210.51$-$080621.3 (hereafter SDSS\,J2232$-$0806) is an AGN at redshift $z=0.276$ (\citealt{Collinson18}).
It was identified as a `slow-blue nuclear hypervariable' object by \cite{Lawrence16} on the basis that it showed large-amplitude optical brightness variability ($|{\Delta g}| \geqslant 1.5$) and the change was slow and blue (occurring over several years, in contrast to the fast and red transients which are likely associated with supernovae).

Our photometric monitoring of this source with the Liverpool Telescope since 2013 has captured one substantial dimming event, and there is sparsely sampled archival photometry that is consistent with similar past events.

\subsection{The aims of this study}
We aim to investigate whether the variability behaviour of this source is best explained by either obscuration of the nucleus, or by some intrinsic change in the emission from the central engine.
The optical spectroscopic monitoring campaign conducted with the William Herschel Telescope allows us to investigate changes in both the AGN continuum and line emission from the broad line region (BLR).

Throughout this paper, we assume a flat $\Lambda$CDM cosmology with $H_0=70$~km~s$^{-1}$~Mpc$^{-1}$, $\Omega_\mathrm{m}=0.3$ and $\Omega_\Lambda=0.7$. 
For the redshift $z=0.276$ this cosmology implies a luminosity distance of 1410.8~Mpc and a flux-to-luminosity conversion factor of $2.38\times10^{56}$~cm$^2$.


\begin{figure*}
\begin{center}
\includegraphics[width=2\columnwidth,keepaspectratio]{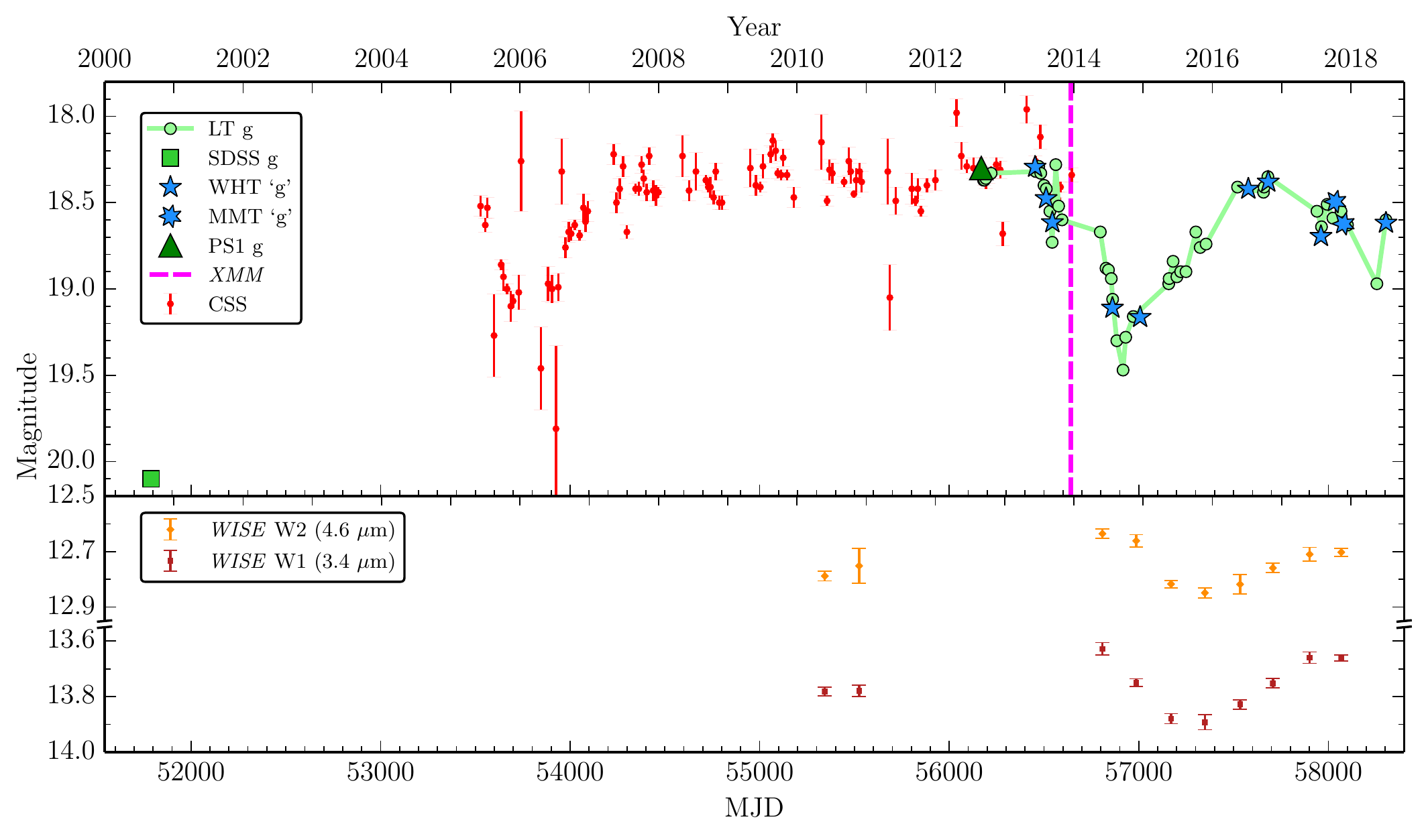}
\end{center}
\caption{\textit{Top}: the optical lightcurve of SDSS\,J2232$-$0806.  The source was observed to brighten by $\Delta g=1.8$~magnitudes between the Sloan Digital Sky Survey (SDSS) observation made in 2000 and the PanSTARRS-1 (PS1) observation of 2012.  We show our follow-up optical photometric monitoring with the Liverpool Telescope (LT) and archival data from the Catalina Sky Survey (CSS).  As well as the direct photometric points, we show the equivalent $g$ magnitudes derived from spectroscopic observations made with the William Herschel Telescope (WHT) and the MMT.  A global greyscale flux correction of $-0.15$~mag has been applied to the spectral magnitudes (see Section~\protect\ref{sec:spec} in the text).  The date of our \textit{XMM-Newton} X-ray and optical-UV observations is also indicated. \textit{Bottom}: the \textit{WISE} infrared lightcurves of SDSS\,J2232$-$0806 (see Section~\protect\ref{sec:wise} in the text).}
\label{fig:lightcurve}
\end{figure*}


\section{The optical monitoring campaign}
\label{sec:data}
\cite{Lawrence16} found that in 2012 the PanSTARRS-1 (PS1) 3$\uppi$ Survey $g$ band photometry of SDSS\,J2232$-0806$ was 1.8 magnitudes brighter than it was in a SDSS photometric observation made in 2000.
To further investigate this interesting source, a photometric monitoring campaign began in 2012 using the Liverpool Telescope and is ongoing.
Optical spectroscopic monitoring commenced in 2013, primarily using the William Herschel Telescope, with an additional two spectra taken in late 2017 with the MMT.
The observing campaign has revealed a dip in brightness of around a factor three in flux and shows a recovery in our most recent observations.
In this section we present our analysis of the optical data.   

\subsection{Observations and data reduction}
\subsubsection{Liverpool Telescope optical photometric monitoring}
\label{sec:lt}
The Liverpool Telescope (LT) is a fully-robotic, remotely controlled 2~m telescope that observes autonomously from La Palma in the Canary Islands.
Photometric observations were taken in the $r$, $g$ and $u$ bands.  
Forty-four independent photometric observations were obtained using the $g$ filter ($\lambda_\mathrm{eff.}=4696$~\AA) between 2012 September and 2018 July are shown in Figure~\ref{fig:lightcurve}.
The $g$ and $r$ bands are much more frequently sampled than the $u$ band, for which we have only twenty-one photometry points.
The observed variability amplitude in the $g$ band ($\Delta g \approx1.2$) is greater than that of $r$ band ($\Delta r \approx0.8$) although we note that the $r$ band ($\lambda_\mathrm{eff}=6111$~\AA) is subject to increasing contamination from the host galaxy as the AGN contribution diminishes.
In addition, the $u$ band ($\lambda_\mathrm{eff}=3499$~\AA) covers the strong, broad Mg~\textsc{ii} emission line (observed at 3573~\AA) and so it is not a clean measure of the AGN continuum. 
For these reasons, in this study we use only photometry obtained in the $g$ band.


\subsubsection{William Herschel Telescope optical spectroscopic monitoring}
\label{sec:wht}
The 4.2~m William Herschel Telescope (WHT) is also situated on the island of La Palma.
SDSS\,J2232$-$0806 has been observed with the WHT on nine occasions between 2013 June and 2018 July.
We used the Intermediate dispersion Spectrograph and Imaging System (ISIS) long-slit, double spectrograph with 
the 5300~\AA\ dichroic which directed the light into the red and blue arms containing the R158B and R300B gratings, respectively.
Typically $\times2$ binning in the spatial direction was used to improve the signal-to-noise ratio (SNR).
This set-up gave a spectral resolution of $R\approx1000$ at 7200~\AA\ in the red and $R\approx1500$ at 5200~\AA\ in the blue, for a slit width of 1~arcsecond.
The total wavelength coverage was $\approx3100$--10600~\AA, this window includes the principal emission lines Mg\,\textsc{ii} $\lambda2800$, H$\upbeta$ $\lambda4861$, [O\,\textsc{iii}] $\lambda\lambda4959,5007$ and H$\upalpha$ $\lambda6563$.

The data reduction was performed with a pipeline using custom \textsc{pyraf} scripts and standard techniques. 
The pipeline is described in detail in \cite{Bruce17} (Section 2.3.3 in that paper). 
  
Unfortunately, we do not have a spectrum contemporaneous with the nadir of the LT lightcurve, which occurred around 2014 September 17.
The spectra obtained on 2014 July 23 and December 16 were recorded 56 days before and 90 days after the photometric minimum and sample the falling and rising side of the dip in the lightcurve, respectively (see Figure~\ref{fig:lightcurve}).  

\subsubsection{MMT spectroscopic monitoring}
\label{sec:mmt}
The MMT is a single 6.5~m mirror telescope 
on Mount Hopkins, Arizona.  
Two optical spectra of SDSS\,J2232$-$0806 were obtained in 2017 December.
The observations were conducted during grey time; on both occasions the observing conditions were clear with sub-arcsecond seeing.
We used the MMT Blue Channel spectrograph with the 300~g~mm$^{-1}$ grating and a 1~arcsecond slit.
This set-up gives a spectral resolution of $R\approx740$ at 4800~\AA, lower than that we obtained with the WHT.  
The target spectra that we use here are the co-added medians of three 10~minute exposures.

\subsection{Optical spectral analysis}
\label{sec:spec}
Optical and ultraviolet fluxes are affected by reddening caused by dust in the Milky Way.
The Galactic neutral hydrogen column density towards SDSS\,J2232$-$0806, $N_\mathrm{H}=4.52\times10^{20}$~cm$^{-2}$ (\citealt{DL90}), implies a colour excess $E(B-V)=0.078$~mag based on the relation derived by \cite{Bohlin78}.
Here, and in Section~\ref{sec:mwl}, we correct our data for Galactic reddening using this value of $E(B-V)$ and the Milky Way reddening curve of \cite{Cardelli89}.

\subsubsection{Internal scaling of spectra}
Before we perform our spectral analysis, we rescale our spectra to account for variations in the absolute flux calibration caused by effects such as seeing (slit losses) and thin cloud.
Since the strong, narrow [O\,\textsc{iii}]~$\lambda5007$ forbidden emission line originates in a low-density, large-volume gas, it should not vary during the course of our monitoring period and is therefore a suitable line to use for internal cross-calibration (provided it it not spatially resolved).  
Rather than simply assuming the flux in the line remains constant (which depends upon an accurate determination of the underlying continuum flux level), we assume instead that the line profile is constant and determine the appropriate flux scaling factors using the \textsc{python} package \textsc{mapspec} developed by \cite{Fausnaugh17}.  
This package is an implementation of, and improvement on, the method of \cite{vanGroningen92}. As noted by them this method should produce a more accurate internal flux scaling than the standard method of simply scaling each spectrum so that the integrated [O\,\textsc{iii}]~$\lambda5007$ line flux is equal to a chosen reference value.

\subsubsection{Absolute flux scaling of spectra}
\label{sec:absflux}
From our internally-scaled optical spectra, we calculated the equivalent LT $g$ magnitude.
The LT optical CCD camera was changed from the RATcam to the IO:O at the end of 2014 February, so in our calculations we use the filter specifications appropriate to the LT instruments in use at the time the spectrum was recorded, although the resultant difference in magnitude is very minor.  
For each spectrum we measured the mean flux $\langle{\nu F_\nu,g}\rangle$ in the LT $g$ band (RATcam 3945--5532~\AA, IO:O 3933--5630~\AA) then calculated the $g$ magnitude equivalent
\begin{equation}
g = -2.5\log\left(\frac{\langle{\nu F_\nu,g}\rangle \times10^{23}}{\nu_\mathrm{eff}~\mathrm{ZP}}\right)~\mathrm{mag}, 
\end{equation}      
where ZP is the zero point magnitude of the filter (RATcam 3940.5~Jy; IO:O 3936.7~Jy) and $\nu_\mathrm{eff}$ is the frequency equivalent to the filter's effective wavelength $\lambda_\mathrm{eff}$ (RATcam 4730~\AA; IO:O 4696~\AA).

By comparison with the LT $g$ magnitudes, we found that the equivalent magnitudes appeared systematically offset by $\approx0.15$~mag.
This slight discrepancy is likely due to slit losses, resulting in a lower flux in our narrow-slit spectra compared with the large-aperture photometry.
Adjusting the magnitudes by $-0.15$~mag (an increase of $\approx15$~per~cent in flux) the equivalent magnitudes replicate both the shape and level of the LT lightcurve, as can be seen in Figure~\ref{fig:lightcurve}.  
In the following, all of the measurements that we make from the spectra include the internal and absolute flux scalings described here.  

\begin{figure*}
\begin{center}
\includegraphics[width=2\columnwidth,keepaspectratio]{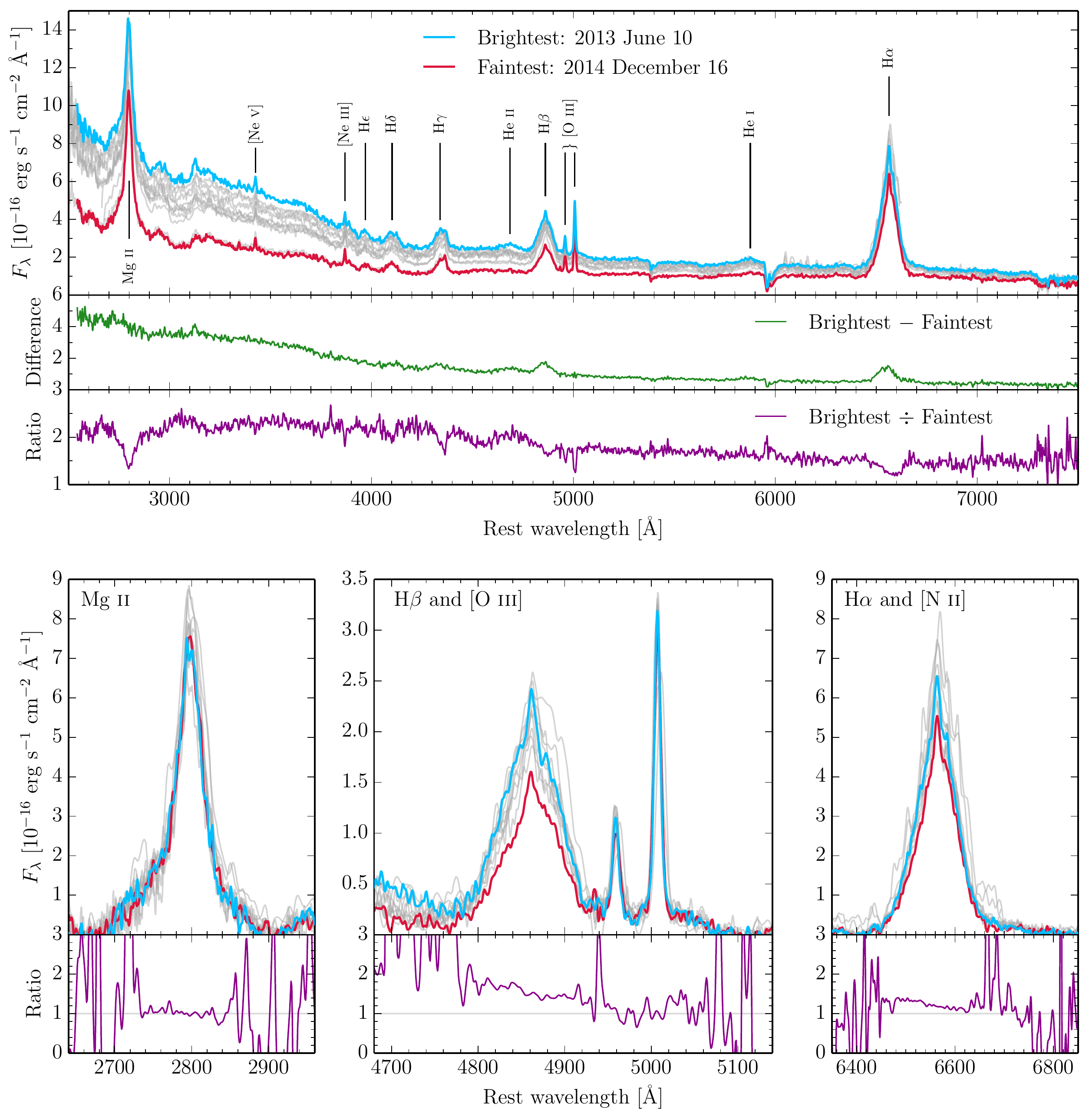}
\end{center}
\caption{\textit{Top}: All eleven optical spectra of SDSS\,J2232$-$0806, rescaled to the same [O\,\textsc{iii}]~$\lambda5007$ emission line profile and corrected for Galactic reddening ($A_V=0.24$).
The brightest spectrum (2013 June) is shown in blue and the faintest spectrum (2014 December) is shown in red; the other nine spectra are shown in grey.  
Prominent emission lines are labelled.
In the lower panels the difference spectrum is shown in green and the ratio spectrum in purple.  
\textit{Bottom}: Continuum-subtracted regions containing key emission lines.  Spectra are colour-coded as in the top plot.
In the lower panels, the ratios between the brightest and faintest spectra is shown in purple.
}
\label{fig:allspec}
\end{figure*}

\subsubsection{Comparison of the optical spectra}
\label{sec:rms}
All eleven optical spectra are shown in the top panel of Figure~\ref{fig:allspec}.
To highlight the spectral variability we have coloured the brightest and faintest spectra in blue and red, respectively, and plotted both their difference and ratio in green and purple, respectively, in the panels below.
The ratio between the brightest and faintest spectrum shows the fractional variability at each wavelength.
The fractional variability at longer wavelengths is diluted by emission from the host galaxy and we see in the ratio spectrum that the fractional variability is greater in the blue end.
Taking the difference removes the constant components including the host galaxy.
In the difference spectrum it can be seen that the absolute flux variation in the blue continuum ($\lambda\lesssim4200$~\AA) is greater than in the red.
The [O\,\textsc{iii}] $\lambda\lambda4959,5007$ lines, which we assumed to be non-variable, are absent in the difference spectrum
which gives us confidence that the flux scaling method we have adopted works well.  
Whereas differences in the H$\upalpha$ and H$\upbeta$ lines between bright and faint spectra are clear, the Mg\,\textsc{ii} line appears to be less variable.
This is obvious in the ratios of the continuum-subtracted lines (shown in purple in the lower panels of the bottom three plots of Figure~\ref{fig:allspec}) where the core of Mg\,\textsc{ii} changes very little and no substantial change is apparent in the broad wings.  
The change in the Balmer lines is most apparent on the blue side of the lines, which seem to have a slight `red shoulder' in the fainter spectra.  
A similar skewness of the H$\upalpha$ profile in the faint state of the CLQ J0159$+$0033 was found by \cite{LaMassa15}.

We computed the mean and root-mean-square (RMS) spectra following the method of \cite{Peterson04}.
Before performing the calculations, the single-epoch spectra are first shifted in wavelength so that the centroids of the [O\,\textsc{iii}] $\lambda5007$ lines (as determined from our model fits) are aligned.
The two MMT spectra are noisier than the nine obtained at the WHT and have less wavelength coverage (particularly redward of H$\upalpha$).
We confirmed that the shapes and general features of our mean and RMS spectra are (broadly) unchanged if we exclude the MMT spectra.
Having done so, we proceeded with the mean and RMS spectra determined from just the WHT observations, so as to extend our results into the red.

\begin{figure}
\begin{center}
\includegraphics[width=\columnwidth,keepaspectratio]{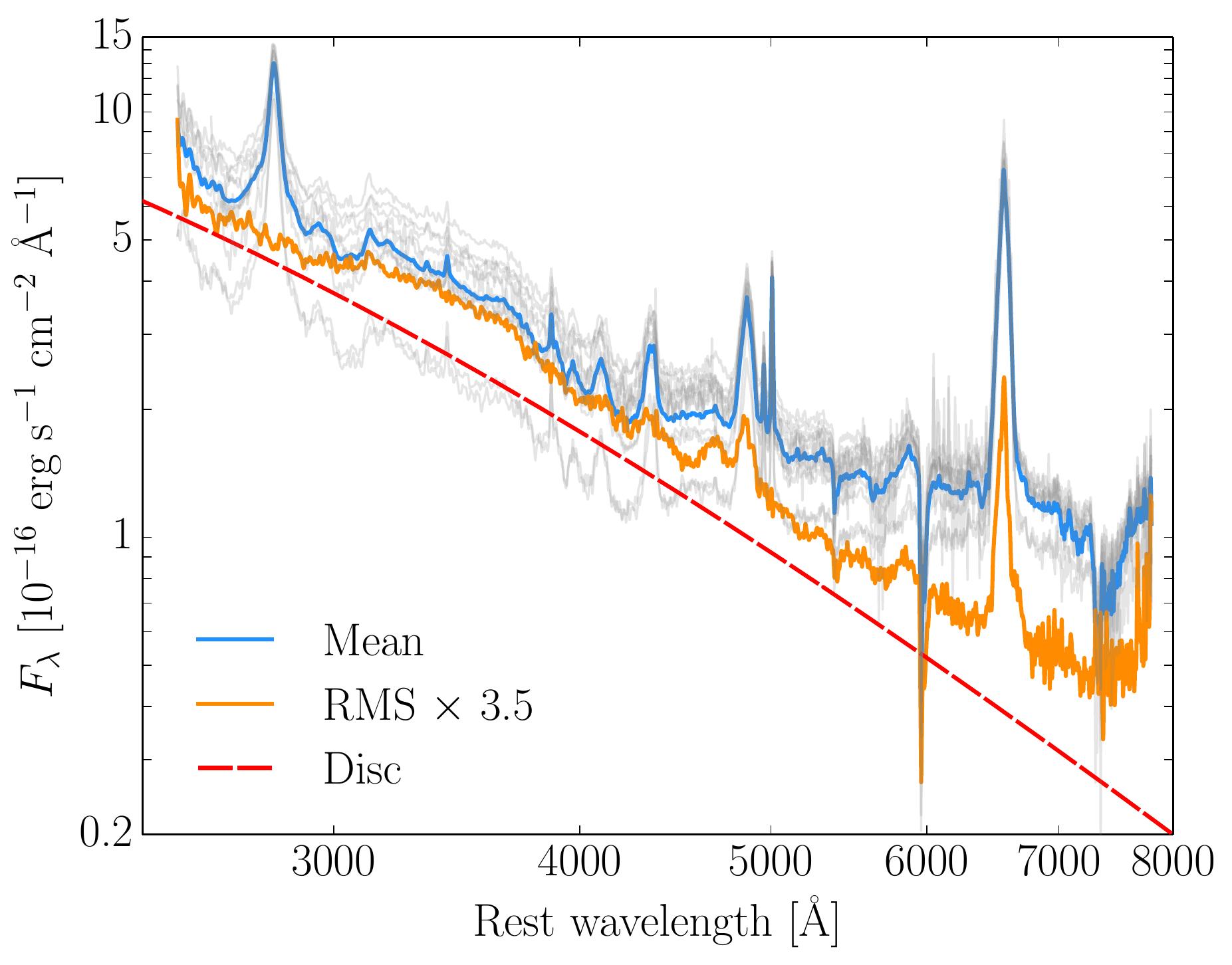} 
\end{center}
\caption{The mean spectrum of SDSS\,J2232$-$0806 is shown in blue and the root-mean-square (RMS) spectrum is shown in orange.
The RMS spectrum has been scaled up by a factor 3.5 to ease the comparison with the mean. 
The eleven spectra from the monitoring campaign are underplotted in grey.
The dashed red line shows the model spectrum of an AGN accretion disc with $L/L_\mathrm{Edd}=3$~per cent and $R_\mathrm{out}=100~R_\mathrm{g}$.   
}
\label{fig:rms}
\end{figure}

The resultant spectra are shown in Figure~\ref{fig:rms}.
As in the difference spectrum, the [O\,\textsc{iii}] lines are removed in the RMS spectrum whereas the Balmer lines, Balmer continuum ($\approx2000$--4000~\AA) and He\,\textsc{ii} $\lambda4685$ are all visible.
We also note that the Mg\,\textsc{ii} emission line is absent from the RMS spectrum, which we discuss later.
Comparing the shapes of the mean and RMS spectra, we see that the RMS spectrum is bluer since the non-variable host galaxy component has been removed: we discuss this in Section~\ref{sec:hostgal}.
The shape of the RMS spectrum is very similar to that of an accretion disc; in Figure~\ref{fig:rms} we show a standard disc spectrum for comparison, calculated for a BH mass of $2\times10^{8}$~M$_{\sun}$, $L/L_\mathrm{Edd}=0.03$ and outer radius of 100 gravitational radii (equal to that determined in our SED model in Section~\ref{sec:optxagnf}).   

\subsection{Measurement of the continuum and the emission lines}
The continuum and emission line fitting was performed using a custom \textsc{python} script employing the \textsc{lmfit} package\footnote{\url{https://lmfit.github.io/lmfit-py/}} which employs a Levenberg-Marquardt algorithm for non-linear, least-squares minimisation. 
The fitting routine appeared to underestimate the errors on the returned parameters, so rather than quoting the error on a single fit, an iterative approach was taken.
Each spectrum was fitted 100 times: on each iteration Gaussian noise was added to the flux density with the amplitude of the noise determined by the measurement error.
The final model parameters and errors are the mean and standard deviation calculated from the 100 iterations.
The standard deviation quantifies the spread of parameter values that can reasonably fit the data.
The errors on the physical quantities derived from the model parameters (e.g.\ the line flux, equivalent width etc.) have been propagated using standard methods.     
The results of our iterative fitting procedure are tabulated in Tables~\ref{tab:WHT1}, \ref{tab:WHT2} and \ref{tab:WHT-MgII} in the Appendix. 

\subsubsection{Red continuum determination}
For the WHT spectra, the (rest frame) 3900--7800~\AA\ continuum is estimated from five emission line free windows of width 50~\AA; these are centred on the wavelengths 4240, 5100, 6205, 7050 and 7700~\AA.  
Because of the narrower wavelength coverage of the two spectra obtained using the MMT, only the first three of these windows are available.  
We fit a power-law continuum of the form $F_\lambda = C\left(\nicefrac{\lambda}{5100~\si{\angstrom}}\right)^{-\alpha}$ through these points to determine the global continuum, allowing the slope $\alpha$ and normalisation $C$ to be free parameters in the fit.  
  
\subsubsection{Modelling of the principal emission lines}
To model the Balmer lines, the red continuum is subtracted from two wavelength windows containing the emission lines of interest (rest frame 4740--5100~\AA\ for H$\upbeta$ and [O\,\textsc{iii}]; 6380--6800~\AA\ for H$\upalpha$ and [N\,\textsc{ii}]).  
The permitted lines were initially fit with a sum of two Gaussians (one broad and one narrow) with the same central wavelength.
However, there were clearly substantial residuals in the line profiles, particularly prominent in the red wing of H$\upalpha$.
We therefore added a third Gaussian component to the Balmer lines, modelling a very broad base, and allowed this to be offset from the central wavelength of the narrower components.  
The two [N\,\textsc{ii}] forbidden lines were each fit with a single, narrow Gaussian.
As well as a strong, narrow Gaussian, a weak, broad Gaussian base was added to the [O\,\textsc{iii}]~$\lambda4959$ and $\lambda5007$ lines.
In all fits we include the permitted Fe~\textsc{ii} emission line template of \cite{Bruhweiler08}, with its normalisation left as a free parameter in the fits.
The model was refined to include the following constraints:  
\begin{enumerate}[i)]  
\item all narrow, broad and very broad lines have the same velocity width (with the exception of the broad bases of the [O\,\textsc{iii}] forbidden lines: these had equal width but this was not tied to the width of the broad permitted lines); 
\item the very broad lines in the H$\upalpha$ and H$\upbeta$ profiles have the same velocity offset;
\item it proved impossible to reliably fit both the width and offset of the very broad lines simultaneously so we fixed the velocity width of these components to $\approx11500~$km~s$^{-1}$ and placed the limit $\Delta v_\mathrm{vb}$~$\lesssim+2500$~km~s$^{-1}$ on the offset\footnote{The line width is approximately equal to the mean FWHM of the very broad Balmer line components of the broad line AGN modelled by \cite{JinWDG12_II}.  The offset was limited to keep the centre of the very broad component within the core of the line.}; 
\item the [O\,\textsc{iii}]~$\lambda4959$ and $\lambda5007$ lines have a fixed flux ratio of 1:3;  
\item the [N\,\textsc{ii}]~$\lambda6548$ and $\lambda6583$ lines have a fixed ratio of 1:3;
\item the stronger [N\,\textsc{ii}]~$\lambda6583$ line has its amplitude fixed to the mean value determined in the WHT spectra;
\item the narrow lines ought not to vary significantly over the monitoring period, therefore the H$\upalpha$ narrow line was fixed to 0.67 of the [O\,\textsc{iii}]~$\lambda5007$ flux, the error-weighted mean value determined from all of the WHT spectra;   
\item the Balmer decrement of the narrow lines was a challenge to determine so was fixed at 6.7, again the error-weighted mean value determined from the WHT spectra.   
\end{enumerate}   
The narrow-line Balmer decrement adopted here is high, although it is within the range $\approx1$--12 found by \cite{JinWDG12_II} for a sample of fifty-one type 1 AGN and at the upper end of the range found by \cite{Lu19} for 554 SDSS DR7 quasars.
If the intrinsic narrow line region (NLR) Balmer decrement is 2.9 (\citealt{O&F06}) the measured value implies an NLR reddening of $A_V\approx2.6$~mag.
However, our aim is to investigate relative changes in the broad line decrement so as long as the subtraction of the narrow line components is consistent, its precise value will have little effect on our results.

\begin{figure*}
\begin{center}
	\begin{tabular}{cc}
	\includegraphics[width=\columnwidth,keepaspectratio]{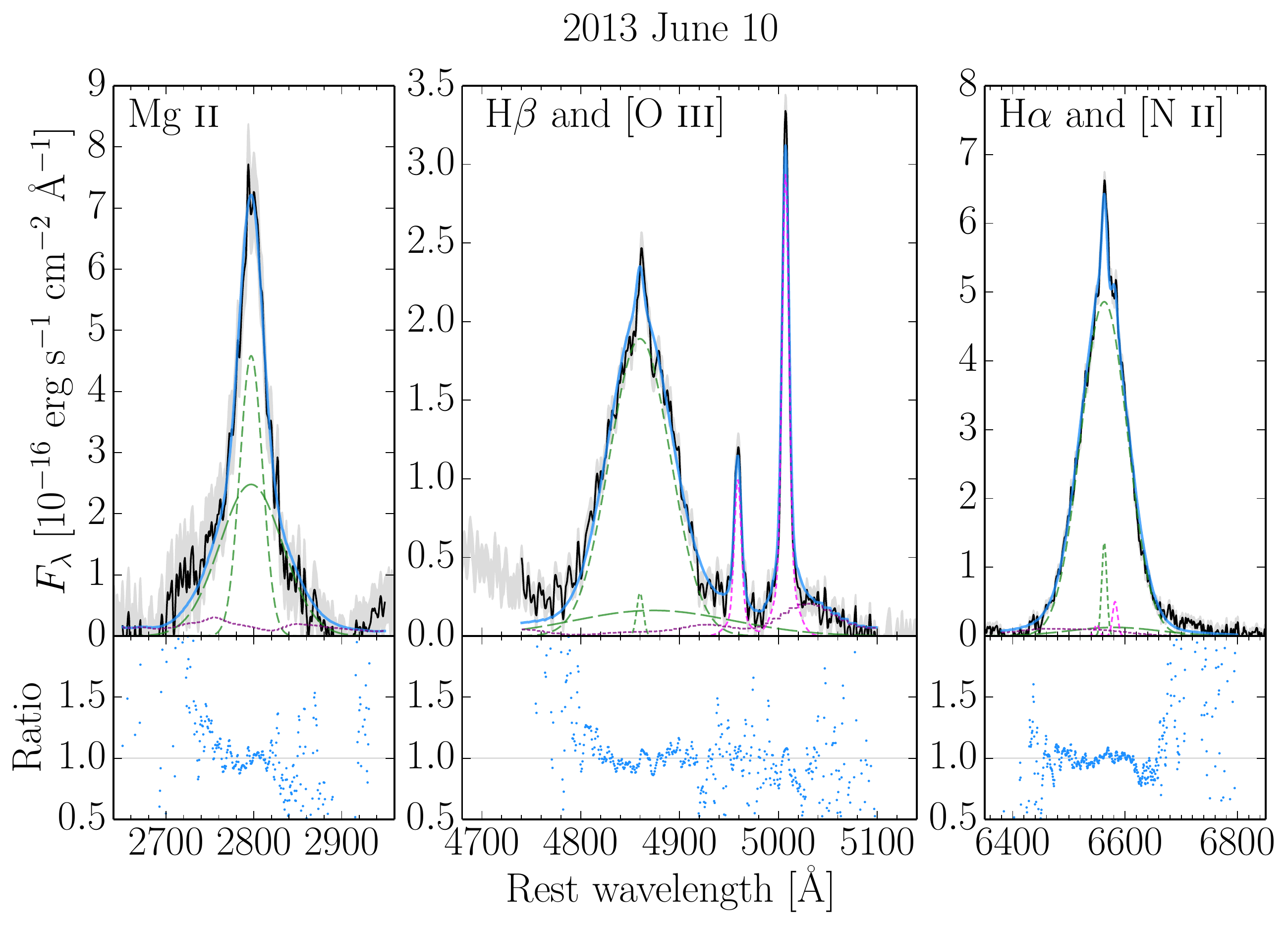} &
	\includegraphics[width=\columnwidth,keepaspectratio]{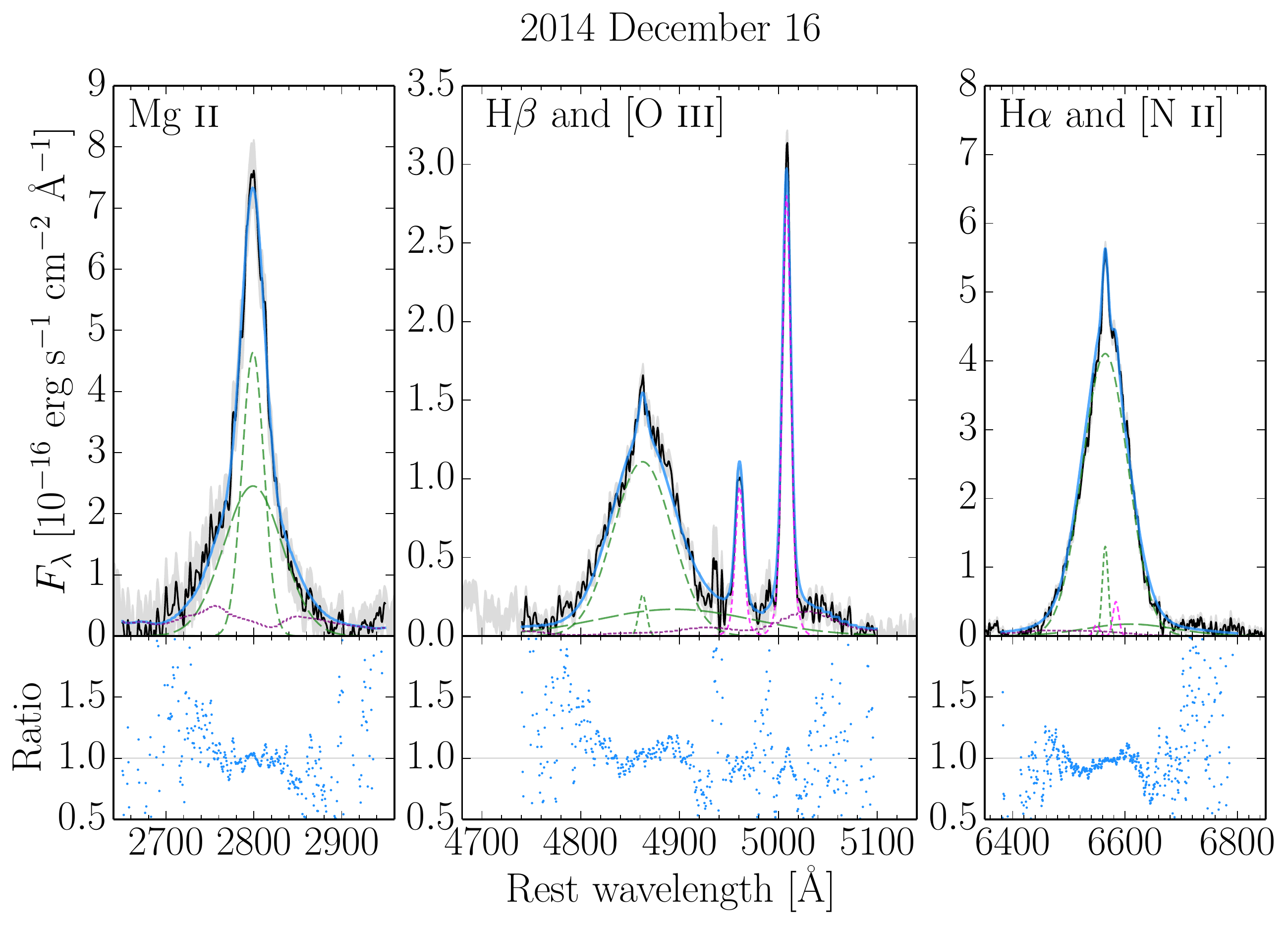} \\
	\end{tabular}
\end{center}
\caption{Examples of emission line fits to continuum-subtracted spectral windows.  In the upper panels, the solid black lines show the wavelength regions of the spectra that were fit, and the solid grey area indicates the error on the flux density; the green short-dashed,  dashed and long-dashed lines show the modelled narrow, broad and very broad components of the permitted lines, respectively; the magenta short-dashed lines show the modelled [O\,\textsc{iii}] and [N\,\textsc{ii}] forbidden lines; an Fe~\textsc{ii} emission template is shown by the purple dotted line and the total model is shown by the solid blue line.  The lower panels show the data / model ratios in the fitted regions.}
\label{fig:linefits}
\end{figure*}

In calculating the Balmer and [O\,\textsc{iii}] emission line EWs we have subtracted the host galaxy contribution to the flux beneath the line (these are determined in \S~\ref{sec:hostgal}) so that the strength of the line is assessed relative to the AGN continuum emission alone\footnote{The host galaxy makes a negligible contribution at the wavelength of Mg\,\textsc{ii}.}.
The Balmer, [O\,\textsc{iii}] and [N\,\textsc{ii}] line properties derived from the best fit model parameters are quoted in Tables~\ref{tab:WHT1} and \ref{tab:WHT2}.
Examples of our Balmer, [O\,\textsc{iii}] and Mg\,\textsc{ii} emission line fits are shown in Figure~\ref{fig:linefits}.

Since there are no emission line free regions in the vicinity of the Mg\,\textsc{ii} line, we do not subtract the continuum before fitting the line.  
Instead we fit the line, Fe~\textsc{ii} template and a power-law continuum simultaneously in the wavelength window 2650--2950~\AA.
The Mg\,\textsc{ii} $\lambda\lambda 2795, 2802$ doublet was not resolved in the composite spectrum produced by stacking the WHT spectra; we therefore fit a single Mg\,\textsc{ii} $\lambda 2800$ profile. 
This emission line was fitted with two Gaussians, one broad and one very broad for the base.
As well as measuring the FWHM of the two components separately, we also calculate the FWHM of the total line profile.
The quantities derived from the best fit model parameters are quoted in Table~\ref{tab:WHT-MgII}.

\subsection{The black hole mass}
\label{sec:mass}
To calculate the mass of the black hole from our emission line and continuum measurements we use the relation
\begin{equation}
M_\mathrm{BH} = K\times L^\alpha \times \mathrm{FWHM}^2
\label{eqn:mass}
\end{equation} 
of \cite{Mejia16} with the appropriate values of $K$ and $\alpha$ taken from their Table 7 (the local calibration corrected for small systematic offsets) for the relevant combinations of the emission line FWHM and continuum or line luminosity $L$.
In Table~\ref{tab:mass} we quote the $K$ and $\alpha$ values used for each relation along with the line and continuum parameters determined as the error-weighted means of values obtained in the four brightest spectra\footnote{Those recorded on 2013 June 10 and August 7 and 2016 July 9 and October 22.}.
We find that the mass is in the range $M_\mathrm{BH}=1.6$--$3.5\times10^{8}$~M$_{\sun}$ (see Table~\ref{tab:mass}), marginally greater than the $1.2$--$1.6\times10^{8}$~M$_{\sun}$ determined by \cite{Collinson18}.
There are considerable uncertainties on the masses estimated by virial methods, which are due to the scatter on the scaling relations.  For relations based on H$\upalpha$ and H$\upbeta$ the 1$\sigma$ scatter is in the range 0.13--0.18~dex; the Mg\,\textsc{ii} relation has a greater scatter of 0.25~dex.  
We adopt a mass of $2\times10^{8}$~M$_{\sun}$ in the following.


\begin{table}
	\centering
	\caption{Black hole mass estimates from optical spectra}
	\label{tab:mass}
	\resizebox{\hsize}{!}{%
	\begin{tabular}{lccccc}
	\hline\Tstrut\Bstrut 
	Relation 												 	& $\log(K)$ & $\alpha$ 	& FWHM 	& $L$  & $M_\mathrm{BH}$\Tstrut\Bstrut  \\
	\hline\Tstrut
	FWHM(H$\upalpha$), $\lambda L_{5100\,\si{\angstrom}}$ 	 	& 6.845 	& 0.650 	& 4.51 	& 1.66 & $2.0$ \\
	FWHM(H$\upalpha$), $L_{\mathrm{H}\upalpha}$ 	 			& 7.389 	& 0.563 	& 4.51 	& 0.13 & $1.6$ \\
	FWHM(H$\upbeta$), $\lambda L_{5100\,\si{\angstrom}}$ 	 	& 6.740 	& 0.650 	& 4.51 	& 1.66 & $1.6$ \\
	FWHM(Mg\,\textsc{ii}), $\lambda L_{3000\,\si{\angstrom}}$ 	& 6.925 	& 0.609 	& 4.19 	& 4.21 & $3.5$\Bstrut   \\
	\hline
	\end{tabular}
	} 
	\parbox[]{\columnwidth}{The broad line FWHMs are in $10^3$~km~s$^{-1}$, the luminosities $L$ in $10^{44}$~erg~s$^{-1}$ and the calculated black hole masses $M_\mathrm{BH} = KL^\alpha \times \mathrm{FWHM}^2$ in $10^8$~M$_{\sun}$.}
\end{table}


\section{Multiwavelength data}
\label{sec:mwl}
\subsection{X-ray and UV observation with \textit{XMM-Newton}}
A 30~ks \textit{XMM-Newton} observation of SDSS\,J2232$-$0806 was made on 2013 December 14 (OBS ID: 0724441001; PI: Lawrence).
At this time, the source was in a relatively high optical flux state (see Figure~\ref{fig:lightcurve}).
The three EPIC X-ray detectors (pn, MOS1, and MOS2) were operating in a Full Frame mode with the Thin filter in place.
Ultraviolet photometry was recorded by the onboard Optical Monitor (OM) which cycled through three of the six filters: U, UVW1 and UVM2.

The data were reduced using the \textit{XMM} Science Analysis Software (\textsc{sas}, v16.0.0) and the latest calibration files available at the time.
The X-ray observation suffered from substantial particle background flaring such that, after filtering, the remaining good time intervals were 8.6, 8.5 and 8.2~ks for the pn, MOS1 and MOS2 detectors, respectively.
The source spectra were extracted from 47~arcsec radius circular regions centred on the source.
The background spectrum was extracted from larger 94~arcsec radius circular regions offset from the source on a blank area of sky.
The spectra were regrouped so as not to oversample the detectors' intrinsic energy resolution by a factor of more than three and to contain at least 20 counts per energy bin, so that they are suitable for a $\chi^2$ analysis.

The OM photometry in the three filters were extracted using the \textsc{sas} tasks \textsc{omichain} and \textsc{omisource}, following the standard procedures.
The OM filter bandpasses cover several emission lines and so do not accurately represent the continuum flux level.
Following the method of \cite{Elvis12}, we can `correct' the photometric fluxes to obtain an improved estimate of the continuum level by multiplying the measured fluxes by the photometric correction factor
\begin{equation}
P_\mathrm{c} = \frac{\mathrm{BW}}{\mathrm{EW}_\mathrm{rest}\times(1+z)+\mathrm{BW}}
\label{eqn:photom-correct}
\end{equation}
where BW is the bandwidth of the photometric filter covering a line of rest-frame equivalent width $\mathrm{EW}_\mathrm{rest}$.
The OM U filter ($\mathrm{BW} = 840$~\AA) covers the Mg\,\textsc{ii} emission line, for which we estimate $\mathrm{EW}_\mathrm{rest}\approx60$~\AA\ $\rightarrow P_\mathrm{c} = 0.92$.
Assuming a C\,\textsc{iii}] $\mathrm{EW}_\mathrm{rest}\approx24$~\AA\ (\citealt{VandenBerk01}), the correction factor in the UVM2 filter is $P_\mathrm{c} = 0.95$.
We conclude that the UVW1 filter is very weakly affected by line emission, since the C\,\textsc{iii}] and Mg\,\textsc{ii} lines only partially appear at the very ends of its bandpass where the sensitivities are lowest. 

\subsubsection{X-ray spectral analysis}
\label{sec:xspec}
Analysis of the X-ray spectra was performed in \textsc{xspec} (\citealt{Arnaud96}) v12.9.1e.
The spectra from the three EPIC detectors were fitted simultaneously, allowing for cross-normalization factors to account for differences in calibration between the detectors; these did not vary by more than 5 per cent.
All models included a Galactic absorption component (\textsc{phabs}) with the column density fixed at $N_\mathrm{H}^\mathrm{Gal}=4.52\times10^{20}$~cm$^{-2}$.
A single power-law was an unsatisfactory fit to the data, giving a reduced $\chi^2$ of 1.27.
A broken power-law was a significant improvement, decreasing the $\chi^2$ value by 48 for the introduction of two additional free parameters and we achieve an acceptable fit with a reduced $\chi^2$ of 1.01.
The $F$-test probability of this improved model was $>99.99$ per cent.
We then tested for an intrinsic absorber by the inclusion of a \textsc{zphabs} component with the redshift fixed to that of the source.
This gave no significant improvement in the fit and we determined an upper limit on the intrinsic column density $N_\mathrm{H}^\mathrm{int}<7\times10^{19}$~cm$^{-2}$.

\begin{table}
	\centering
	\caption{X-ray spectral models}
	\label{tab:xmm}
	\begin{tabular}{llc}
	\hline 
	Model 						& Parameter 											& Value \\
	\hline
	\Tstrut\Bstrut  
	\textsc{powerlaw}			& $\Gamma$  											& $2.19\pm0.02$  \\
								& Norm.													& $(2.83\pm0.04)\times10^{-4}$ \\
								& $\chi^2$/d.o.f.										& $223/175=1.27$ \\
	\Tstrut
	\textsc{bknpower}   		& $\Gamma_1$  											& $2.35^{+0.04}_{-0.05}$ \\
								& $E_\mathrm{brk}$ (keV)  								& $1.7^{+0.5}_{-0.2}$ \\
								& $\Gamma_2$  											& $1.79^{+0.07}_{-0.17}$ \\
								& Norm.													& $\left(2.71^{+0.07}_{-0.05}\right)\times10^{-4}$ \\
								& $\chi^2$/d.o.f.										& $175/173=1.01$ \\
	\Tstrut
	\textsc{zphabs}~$\times$    & $N_\mathrm{H}^\mathrm{int}$ ($10^{19}$~cm$^{-2}$)		& $<7$ \\
	\textsc{bknpower}   		& $\Gamma_1$  											& $2.35^{+0.04}_{-0.02}$ \\
								& $E_\mathrm{brk}$ (keV)  								& $1.7^{+0.5}_{-0.2}$ \\
								& $\Gamma_2$  											& $1.79^{+0.07}_{-0.20}$ \\
								& Norm.													& $\left(2.71^{+0.06}_{-0.05}\right)\times10^{-4}$ \\
								& $\chi^2$/d.o.f.										& $175/172=1.01$ \\
	\hline 		
	\end{tabular}
	\parbox[]{7.2cm}{All models included a Galactic absorption component (\textsc{phabs}) with the column density fixed at $N_\mathrm{H}^\mathrm{Gal}=4.52\times10^{20}$~cm$^{-2}$.}		
\end{table} 

\subsection{Archival photometric data}
\subsubsection{\textit{WISE}}
\label{sec:wise}
The \textit{Wide-field Infrared Survey Explorer} (\textit{WISE}, \citealt{WISE10}) telescope observed SDSS\,J2232$-$0806 twice in 2010.
Data for this source was found in the AllWISE Source Catalog, hosted by the Infrared Science Archive (IRSA\footnote{\url{http://irsa.ipac.caltech.edu/}});  
in Table~\ref{tab:mwl} we quote the reported instrumental profile-fit magnitudes.
The photometric quality of these detections were A (best) for the W1, W2 and W3 filters and B for the W4 filter. 

As well as the catalogue magnitudes, we also obtained infrared lightcurves in the W1 and W2 filters from the \textit{WISE} and \textit{Near-Earth Orbit WISE Reactivation} (\textit{NEOWISE}) archives\footnote{Because of the depletion of hydrogen coolant, only the W1 and W2 filters have been operable since the beginning of the \textit{NEOWISE} mission.}.
In addition to the two visits made during the \textit{WISE} mission, SDSS\,J2232$-$0806 has been observed with roughly six-month cadence since the start of \textit{NEOWISE} mission in December 2013.
Typically a dozen exposures are made on each visit; to construct the lightcurves shown in Figure~\ref{fig:lightcurve}, we have calculated the mean and standard error on the magnitudes recorded on each visit.
We exclude the seven exposures taken on MJD 57345, because there was a large scatter on these magnitudes and a set of eleven exposures was taken three days later.
This visit on MJD 57348 (2015 November 19) corresponds to the minima of the infrared lightcurves and occurs 428 days later than the observed minimum in the LT optical lightcurve (see Section~\ref{sec:lt}).
There is a 0.26~mag peak-to-trough change in W1 and a 0.21~mag change in W2. 

\subsubsection{Two-Micron All Sky Survey}
SDSS\,J2232$-$0806 was observed as part of the Two-Micron All Sky Survey (2MASS, \citealt{Skrutskie06}), which was conducted between 1997 and 2001.
In Table~\ref{tab:mwl} we quote the $J$, $H$ and $K_s$ profile-fit magnitudes reported in the 2MASS All-Sky Point Source Catalog (PSC)\footnote{Also available from IRSA, see earlier note.}.
The observation was made on 1998 October 1 and the photometric quality flag is C for all filters.

\subsubsection{Sloan Digital Sky Survey}
Although no Sloan Digital Sky Survey (SDSS) spectroscopic data exists for this source, photometry was obtained on 2000 March 9.
As can been seen in Figure~\ref{fig:lightcurve}, the source was in a very low state at this time.
The object was classified as a (passive) galaxy based on its photometric colours.

\subsubsection{PanSTARRS-1 3$\pi$ Survey}
The PanSTARRS-1 (PS1) 3$\uppi$ Survey was conducted between 2009 and 2014, observing the $\nicefrac{3}{4}$ of the sky north of $-30^{\circ}$ declination multiple times per year in each of five filters (see \citealt{Magnier13} and \citealt{Chambers16}).  
Originally searching for tidal disruption events, \cite{Lawrence16} identified SDSS\,J2232$-$0806\footnote{The common name of the source in this paper is J223210.} as one of a number of `slow blue nuclear hypervariables': objects with no previously known AGN, blue colours and evolution on timescales of years.  
This particular source was brighter by $\Delta g = 1.80\pm0.04$ in 2012 compared with the SDSS photometry of 2000.

\subsubsection{UK Schmidt Telescope}
We located a record for SDSS\,J2232$-$0806 in the SuperCOSMOS Science Archive (SSA\footnote{\url{http://ssa.roe.ac.uk/}.}).
The $B_j$ band ($\lambda=3950$~\AA) observation was made using the UK Schmidt Telescope (\citealt{Canon75}) on Siding Spring Mountain, NSW, Australia, on 1986 August 1.
Its sCorMag (stellar magnitude in the Vega system) is given in the SSA as $B_j=19.02$~mag. 
Converting this to a $g$ band AB magnitude, we estimate $g\approx18.6\pm0.3$, where the uncertainty is the standard single-passband uncertainty on SuperCOSMOS magnitudes (\citealt{Hambly01}).

\subsubsection{Hubble Space Telescope}
\label{sec:HST}
Two short-exposure photometric observations were made with the Wide Field Camera 3 (WFC3) onboard the \textit{Hubble Space Telescope} (\textit{HST}) on 2015 September 18.
The exposure times were 330~s in the wide IR F125W filter ($\lambda_\mathrm{eff}=1.25$~$\upmu$m, $J$ band) and 1200~s in the extremely wide UVIS F475X filter ($\lambda_\mathrm{eff}\approx4776$~\AA, and including the $g$ band).
     
\subsubsection{\textit{GALEX}}
Two epochs of ultraviolet (UV) photometry were found by searching the \textit{Galaxy Evolution Explorer} (\textit{GALEX}, \citealt{Galex05}) space telescope archive.
In both records, the UV source is coincident with the optical coordinates of SDSS\,J2232$-$0806 within 1.3~arcsec.
The UV flux increases by a factor $\approx3$ between the two epochs and there is also an apparent colour change, with SDSS\,J2232$-$0806 appearing bluer in the later observation. 

\begin{table*}
	\centering
	\caption{The multiwavelength photometric dataset}
	\label{tab:mwl}
	\begin{tabular}{lllccccc}
	\hline 
	Date 				& Telescope or 				& Filter	& Measurement 		& Unit 			& $\log(\nu)^\mathrm{a}$	& Flux$^\mathrm{b}$ 		& Luminosity$^\mathrm{c}$ \\
						& survey					&			&					&				& 							& 							&  \\
	\hline
	2010/05/27--28		& \textit{WISE}				& W4		& $7.536\pm0.157$	& Vega mag		& 13.13						& $1.10\pm0.16$				& $2.62\pm0.38$ \\ 
	2010/05/27--28		& \textit{WISE}				& W3		& $10.258\pm0.073$	& Vega mag		& 13.41						& $6.48\pm0.44$				& $15.4\pm1.0$ \\
	2010/05/27--11/25	& \textit{WISE}				& W2		& $12.783\pm0.027$	& Vega mag		& 13.81						& $8.62\pm0.21$				& $20.5\pm0.5$ \\
	2010/05/27--11/25	& \textit{WISE}				& W1		& $13.782\pm0.027$	& Vega mag		& 13.95						& $8.50\pm0.21$				& $20.2\pm0.5$ \\ 
	1998/10/01			& 2MASS						& $K_s$		& $15.419\pm0.182$	& mag			& 14.14						& $6.29\pm1.06$				& $15.0\pm2.5$ \\
	1998/10/01			& 2MASS						& $H$		& $16.166\pm0.208$	& mag			& 14.26						& $6.31\pm1.21$				& $15.0\pm2.9$ \\
	1998/10/01			& 2MASS						& $J$		& $17.124\pm0.201$	& mag			& 14.39						& $5.47\pm1.01$				& $13.0\pm2.4$ \\
	2000/03/09			& SDSS						& $z$		& $18.33\pm0.04$	& asinh mag		& 14.53						& $5.6\pm0.2$				& $14.8\pm0.5$	\\
	2000/03/09			& SDSS						& $i$		& $18.97\pm0.02$	& asinh mag		& 14.61						& $3.78\pm0.07$				& $10.4\pm0.2$	\\
	2000/03/09			& SDSS						& $r$		& $19.20\pm0.02$	& asinh mag		& 14.69						& $3.71\pm0.07$				& $10.8\pm0.2$	\\
	2000/03/09			& SDSS						& $g$		& $20.10\pm0.03$	& asinh mag		& 14.81						& $2.14\pm0.06$				& $6.7\pm0.2$	\\
	2012/08/30			& PanSTARRS-1 3$\uppi$		& $g$		& $18.30\pm0.05$	& mag			& 14.80						& $11.0\pm0.5$				& $34\pm2$	\\
	2012/09/11			& Liverpool					& $g$		& $18.37\pm0.02$	& AB mag		& 14.81						& $10.5\pm0.2$				& $33.0\pm0.6$ \\
	1988/06/01	       	& Schmidt					& $g^*$		& $18.6\pm0.3$		& AB mag		& 14.81						& $9\pm2$					& $27\pm6$ \\
	1988/06/01	       	& Schmidt					& $B_j$		& $19.0\pm0.3$		& Vega mag		& 14.88						& $10\pm3$					& $33\pm7$ \\
	2000/03/09			& SDSS						& $u$		& $19.79\pm0.05$	& asinh mag		& 14.92						& $3.8\pm0.2$				& $12.7\pm0.7$	\\
	2013/12/14			& \textit{XMM-Newton} OM	& U			& $2.219\pm0.038$	& cts~s$^{-1}$	& 14.94						& $14.8\pm0.2$				& $50.4\pm0.7$	\\
	2013/12/14			& \textit{XMM-Newton} OM	& UVW1		& $1.154\pm0.017$	& cts~s$^{-1}$	& 15.01						& $16.2\pm0.2$				& $58.5\pm0.7$ \\
	2013/12/14			& \textit{XMM-Newton} OM	& UVM2		& $0.390\pm0.011$	& cts~s$^{-1}$	& 15.11						& $19.9\pm0.6$				& $89\pm3$	\\
	2003/08/22			& \textit{GALEX} 			& NUV		& $20.67\pm0.22$	& AB mag		& 15.12						& $2.6\pm0.5$				& $12\pm2$  \\
	2004/08/24			& \textit{GALEX}			& NUV		& $19.81\pm0.04$	& AB mag		& 15.12						& $6.8\pm0.2$				& $31\pm1$ \\
	2003/08/22			& \textit{GALEX}			& FUV		& $21.10\pm0.32$	& AB mag		& 15.29						& $2.6\pm0.7$				& $11\pm3$	\\
	2004/08/24			& \textit{GALEX}			& FUV		& $19.62\pm0.03$	& AB mag		& 15.29						& $8.4\pm0.3$				& $36\pm1$	\\
	\hline 		
	\end{tabular}
	\parbox[]{16cm}{\textit{Notes}: $^\mathrm{a}$Logarithm of the observed frequency $\nu$ in Hz; $^\mathrm{b}$observed flux $\nu F_\nu$ in units of $10^{-13}$~erg~s$^{-1}$~cm$^{-2}$; $^\mathrm{c}$intrinsic luminosity $\nu L_\nu$ in units of $10^{43}$~erg~s$^{-1}$, dereddened where appropriate. $^*$Converted from the quoted $B_j$ magnitude below.}		
\end{table*}

\subsection{The spectral energy distribution}
\subsubsection{Host galaxy contribution to the SED and spectra}
\label{sec:hostgal}
Infrared and optical emission from the host galaxy bulge may make a non-negligible contribution to our spectra, particularly in the faint state.
It can be seen in our SED (Figure~\ref{fig:sed}) that the bulge component 
dominates over the AGN continuum redward of H$\upbeta$.
However, this is not representative of the host galaxy flux in our spectra, since our narrow 1~arcsec wide slit excludes much of the extended host galaxy emission: a typical bulge diameter of 15~kpc would be $\approx3.6$~arcsecs across on the sky.

We examined the \textit{HST} images of the source, taken in 2015 September (see Section~\ref{sec:HST}).
The high spatial resolution of the instrument in principle allows us to separate the point-like AGN emission from the more extended host galaxy.
We made a visual inspection of the 1D brightness profiles of the source in the two filters.
Whereas the source emission in the UVIS filter was PSF-like, the $J$ band profile had a slightly more extended base than the PSF, suggesting the presence of some light from the host galaxy.
Unfortunately, however, the snapshot \textit{HST} exposures are not sufficiently deep to robustly assess the host galaxy emission.

Instead, we can estimate the host galaxy luminosity at 5100~\AA\ in our spectral extraction aperture using the relation of \cite{Landt11}.
From a sample of low-redshift ($z\lesssim0.3$), bright, broad emission line AGN, the authors determined the host galaxy luminosities enclosed in the apertures from stacked \textit{HST} images (see their Section 3 and Figure 1).
When extracting the WHT spectra, we integrated on average 4.75~arcsec in the spatial direction; the 4.75~arcsec$^2$ aperture is therefore equivalent to a spatial size of 20~kpc$^2$ at the source.
From the \cite{Landt11} relation we then estimate $F_{5100\si\angstrom}\approx4.2\times10^{-17}$~erg~s$^{-1}$~cm$^{-2}$~\AA$^{-1}$.







The RMS spectrum we constructed in Section~\ref{sec:rms} largely removes the non-variable host galaxy contribution, whereas the mean spectrum does not.
Therefore, if we assume that the mean AGN emission has the same spectral shape as the variable component, we can estimate the host galaxy contribution by the `red excess' of the mean spectrum in comparison with the RMS.
For the host galaxy component we used the 5~Gyr old elliptical galaxy template of \cite{Polletta07}.   
We add the RMS and host galaxy spectra, and rescale the two components until the sum satisfactorily matches the shape of the mean spectrum.  
From the appropriately-scaled galaxy template we determine the mean flux densities in several 150~\AA\ wide windows.
The flux densities at 4861, 5007, 5100 and 6563~\AA\ are 4.9, 4.8, 4.6 and $4.8 \times 10^{-17}$~erg~s$^{-1}$~cm$^{-2}$~\AA$^{-1}$, respectively; the value at 5100~\AA\ is consistent with the \cite{Landt11} estimate calculated above, given the uncertainties.
The host galaxy contribution to the fluxes at 2800~\AA\ (under Mg\,\textsc{ii}) and at 3000~\AA\ is negligible.
In the rest of this study we correct the AGN continuum fluxes (and hence the emission line EWs) using these values.
The emission line EWs recorded in the Tables in the Appendix reflect this correction.

\subsubsection{Accretion flow model}
\label{sec:optxagnf}
To model the multiwavelength SED, we use the energy-conserving accretion flow model \textsc{optxagnf} of \cite{Done12}.  
The model has a standard thin accretion disc from outer radius $R_\mathrm{out}$ to $R_\mathrm{cor}$.
Interior to $R_\mathrm{cor}$, the accretion power is divided between soft and hard Comptonisation regions.
The hard Comptonisation region receives the fraction $f_\mathrm{pl}$ of the available accretion power and produces power-law emission with photon index $\Gamma$.
The soft Comptonisation region upscatters seed photons from the inner edge of the standard thin disc producing soft X-ray emission in excess of the hard coronal power-law (this emission is often called the `soft X-ray excess': SX). 
The soft Comptonisation region is parameterised by its optical depth $\tau$ and warm electron temperature $kT_\mathrm{e}$. 

\begin{figure*}
\begin{center}
\includegraphics[width=2\columnwidth,keepaspectratio]{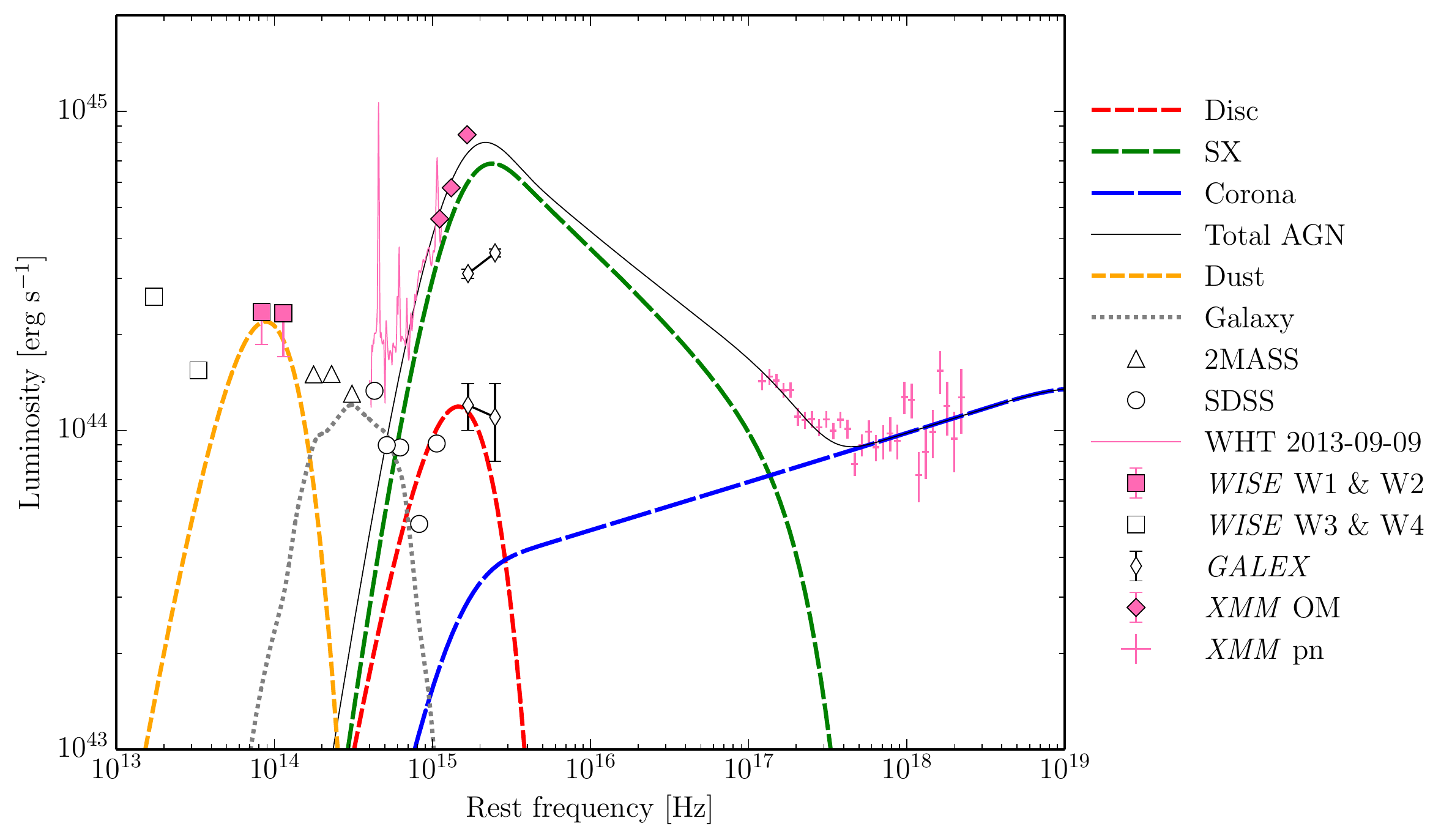} 
\end{center}
\caption{The multiwavelength spectral energy distribution of SDSS\,J2232$-$0806.
Modelled data are shown in pink: the \textit{XMM-Newton} OM and EPIC-pn data of 2013 December 14; WHT spectrum of 2013 September 9 and \textit{WISE} W1 and W2 IR photometry.
Additionally, we show other archival data in white: \textit{WISE} W3 and W4 IR photometry from 2010; 2MASS IR photometry from 1998; SDSS photometry from 2000 and two epochs of \textit{GALEX} UV photometry from 2003 (faint) and 2004 (bright).}
\label{fig:sed}
\end{figure*}

In addition to the direct accretion flow emission, we include a redshifted blackbody (\textsc{zbbody}) modelling the hot dust which is sampled by the \textit{WISE} W1 and W2 bands.
In Figure~\ref{fig:sed} we show the W1 and W2 fluxes corresponding to the earliest \textit{NEOWISE} observation (2014 May 31: the closest in time to the \textit{XMM-Newton} pointing). 
The downward error bars show the extent of the flux diminution over the observing period.
For completeness, the figure also shows the \textit{WISE} W3 and W4 band fluxes, which sample cooler dust.
We do not model these data points; the emission may be attributed to AGN- or starlight-heated dust (or some mixture of the two). 
We show our model SED in Figure~\ref{fig:sed}, along with the modelled multiwavelength data.  
Archival data are also shown for illustrative purposes, including two epochs of \textit{GALEX} UV photometry, 2MASS infrared photometry and the SDSS optical photometry from 2000 during which the AGN was in a deep flux minimum.
In Figure~\ref{fig:sed} we also show the \cite{Polletta07} 5~Gyr old elliptical host galaxy template which is normalised to fit the SDSS photometry.

Our SED model has a very prominent soft Comptonisation region that emits from the optical/UV into the soft X-ray band.
The standard disc component is required only to provide a source of seed photons for the soft Comptonisation region in the model calculations and not to fit the shape of the SED itself.
We note that \cite{Collinson18} presented an alternative SED model which contained no soft Comptonisation region and in which the optical/UV emission was attributed to a standard accretion disc, with the X-ray spectrum modelled by a single power-law component.
This model cannot replicate the curvature in the X-ray spectrum which we detected significantly in Section~\ref{sec:xspec}.
Additionally, whilst the single power-law of \cite{Collinson18} has a photon index of $\Gamma=2.2$, a harder index (such as the $\Gamma=1.85$ we determine here) would be expected for a system of this Eddington ratio (e.g.\ \citealt{Kubota18}).
However, the Eddington ratio determined in both models, $\nicefrac{L}{L_\mathrm{Edd}}=0.1$, is the same.  

\begin{table*}
	\centering
	\caption{Multiwavelength SED model parameters}
	\label{tab:sed}
	\begin{tabular}{llllc}
	\hline
	Model 				& Parameter  							& Units 					& Description 														& Value \\
	\hline
	%
	%
	\textsc{zbbody}		& $kT_\mathrm{dust}$					& keV (K)					& Hot dust temperature												& $9.30\times10^{-5}$ ($1140$) \\
						& B.body norm.							& 							& Hot dust blackbody normalisation									& $2.43\times10^{-5}$ \\
	\textsc{hostpol}	& Gal. norm. 							& 	   						& Host galaxy template normalisation 								& $2.71\times10^{-7}$	\\
	\textsc{optxagnf}	& $\log(\nicefrac{L}{L_\mathrm{Edd}})$	&							& Eddington ratio													& $-1.00$  \\
						& $kT_\mathrm{e}$						& keV  						& Electron temperature of soft Comptonisation region				& $0.20$  \\
						& $\tau$								&   						& Optical depth of soft Comptonisation region						& $17.3$  \\
						& $\Gamma$								&           				& Photon index of power-law coronal emission						& $1.85$  \\
						& $f_\mathrm{pl}$						&							& Fraction of power below $R_\mathrm{cor}$ emitted in power-law		& $0.32$  \\
						& $R_\mathrm{cor}$						& $R_\mathrm{g}$			& Inner (standard) accretion disc radius							& $80.0$  \\
						& $\log(R_\mathrm{out})$				& $R_\mathrm{g}$			& Outer accretion disc radius										& $2.01$ \\
						& $F_\mathrm{dust}$						& erg~s$^{-1}$~cm$^{-2}$	& Flux of hot dust blackbody										& $1.25\times10^{-12}$ \\
						& $F_\mathrm{disc}$						& erg~s$^{-1}$~cm$^{-2}$	& Flux of (standard) accretion disc									& $6.81\times10^{-13}$ \\
						& $F_\mathrm{SX}$						& erg~s$^{-1}$~cm$^{-2}$	& Flux of of soft Compton emission									& $8.03\times10^{-12}$ \\
						& $F_\mathrm{pl}$						& erg~s$^{-1}$~cm$^{-2}$	& Flux of coronal power-law emission								& $3.78\times10^{-12}$ \\
						& $F_\mathrm{UV}$						& erg~s$^{-1}$~cm$^{-2}$	& AGN flux between 100--4000~\AA\ (rest-frame)						& $7.92\times10^{-12}$ \\
						& $F_\mathrm{AGN}$						& erg~s$^{-1}$~cm$^{-2}$	& Total AGN flux													& $1.25\times10^{-11}$ \\
	\hline
	\end{tabular}
	\parbox[]{16cm}{\textit{Note:} Distances are measured in gravitational radii $R_\mathrm{g}=GM_\mathrm{BH}/c^2$.}		
\end{table*}

\section{The nature of the variability} 
We now bring together all of these data-sets, and use them to confront two generically distinct scenarios i.e.\ that the flux changes seen in Figure~\ref{fig:allspec} are due to reddening by dust, or, that they are a result of an intrinsic variation in the continuum emission from the nuclear region, primarily powered by processes occurring within the accretion disc. 

\subsection{Obscuration interpretation}
\begin{figure}
\begin{center}
\includegraphics[width=\columnwidth,keepaspectratio]{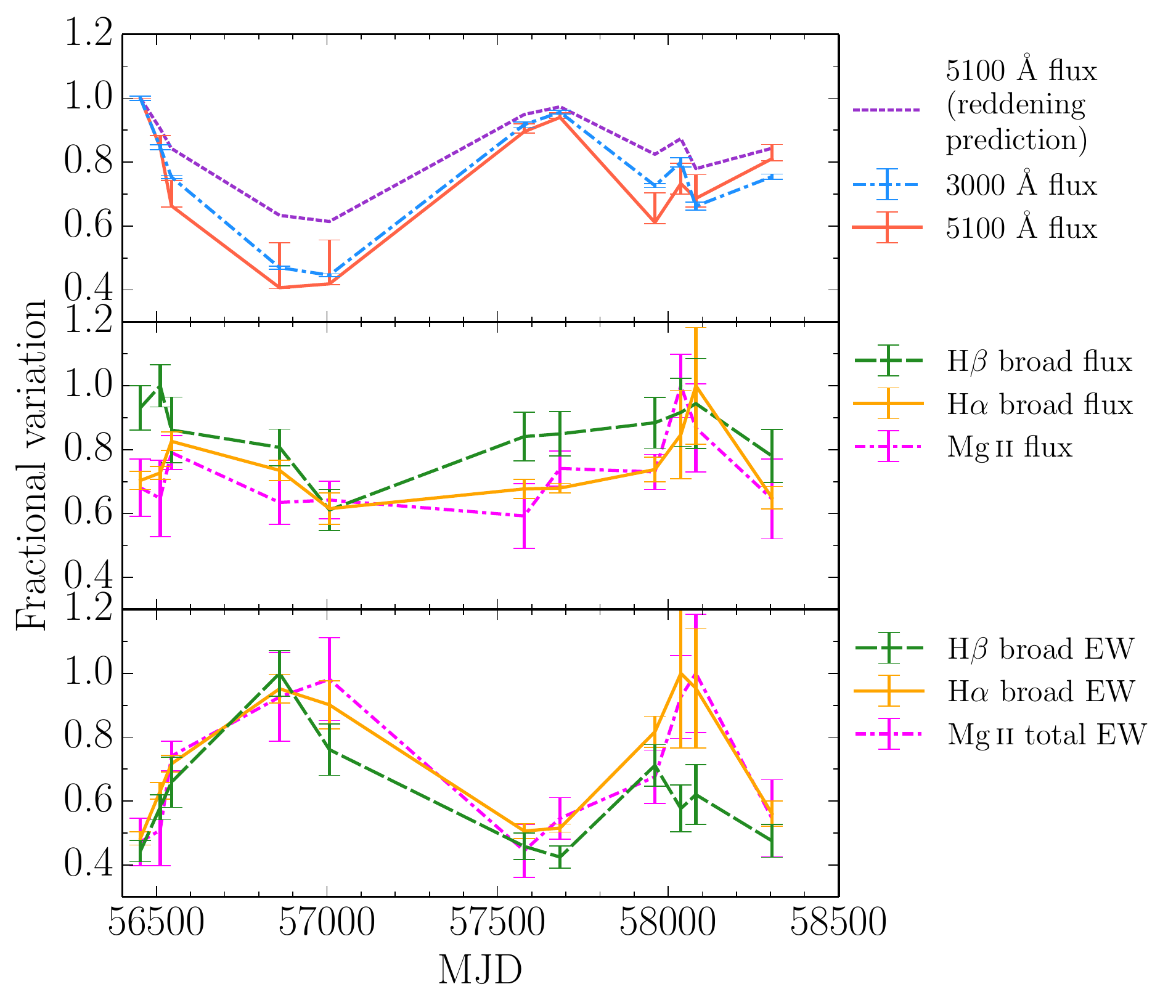} 
\end{center}
\caption{Fractional variations in the 3000 and 5100~\AA\ continuum fluxes and emission line fluxes and equivalent widths (EWs) over the monitoring period. 
In the top panel, as well as the observed continuum variations we also show the predicted 5100 \AA\ variations, calculated from the observed 3000 \AA\ variations, on the assumption that these are caused by reddening (see text).
The measured 5100~\AA\ fluxes have been corrected for host galaxy contamination using our estimate determined in Section~\protect\ref{sec:hostgal} in the text; the upper error bars indicate the fractional variations calculated with no host galaxy subtraction.
For the Balmer lines, values are calculated from the sum of very broad and broad components; the Mg\,\textsc{ii} values are calculated from the whole line profile.}
\label{fig:var}
\end{figure}

In Figure~\ref{fig:var} we show the relative variations of the continuum fluxes and those of the Mg\,\textsc{ii} and broad Balmer emission lines.
The estimated host galaxy flux at 5100~\AA\ has been subtracted from the red continuum flux (see Section~\ref{sec:hostgal}).
Shorter wavelengths are more sensitive to reddening than longer ones so, under the assumption that the observed changes are due to reddening, we would expect the 5100~\AA\ flux to have a shallower fractional variability curve than at 3000~\AA.
Based on the \cite{Cardelli89} Milky Way\footnote{We note the reddening curves for the Small and Large Magellanic Clouds are very similar to the Milky Way curve for wavelengths $>3000$~\AA\ which we consider here.} reddening curve, we have calculated the extinction ($A_V$) required to cause the observed fractional changes in the blue continuum and then predict the fractional change in the red continuum for the same $A_V$.
We see that the observed AGN flux at 5100~\AA\ shows a significantly greater fractional variability than this prediction, and is broadly consistent with the fractional variations at 3000~\AA. 
There is considerable uncertainty in the AGN continuum 5100~\AA\ fractional flux variations due to the uncertainty in the host galaxy flux subtraction. 
However, even in the very conservative case when we perform no host galaxy flux subtraction, the 5100~\AA\ fractional flux variability is still inconsistent with that predicted from a reddening law (as indicated by the upper error bars in Figure~\ref{fig:var}).

Additionally, we find that the amplitude of line flux changes are somewhat lower than those in the continuum.
The 3000~\AA\ flux exhibits variability of more than a factor two whereas the lines show only $\approx40$~per cent decrease.
(Note that our spectroscopic observations did not cover the deep flux minimum seen in the photometric lightcurve.)
There is a trend for the emission line EWs to be anticorrellated with the continuum fluxes: increasing when the continuum dims and vice versa. 
We find that the minimum (maximum) emission line EWs determined over the spectroscopic monitoring period are 570 (1200), 110 (250) and 50 (110)~\AA\ for H$\upalpha$, H$\upbeta$ and Mg\,\textsc{ii}, respectively.
In the case of a simple screen obscuring both the accretion disc (from which the continuum originates) and BLR (from which the broad lines originate), the equivalent widths of the lines ought not to change since both continuum and line flux at any given wavelength will be suppressed equally.
However, if the absorber covers more of the very compact accretion disc than the larger BLR then the EW of the broad lines would be seen to increase. 

\begin{figure}
\begin{center}
\includegraphics[width=\columnwidth,keepaspectratio]{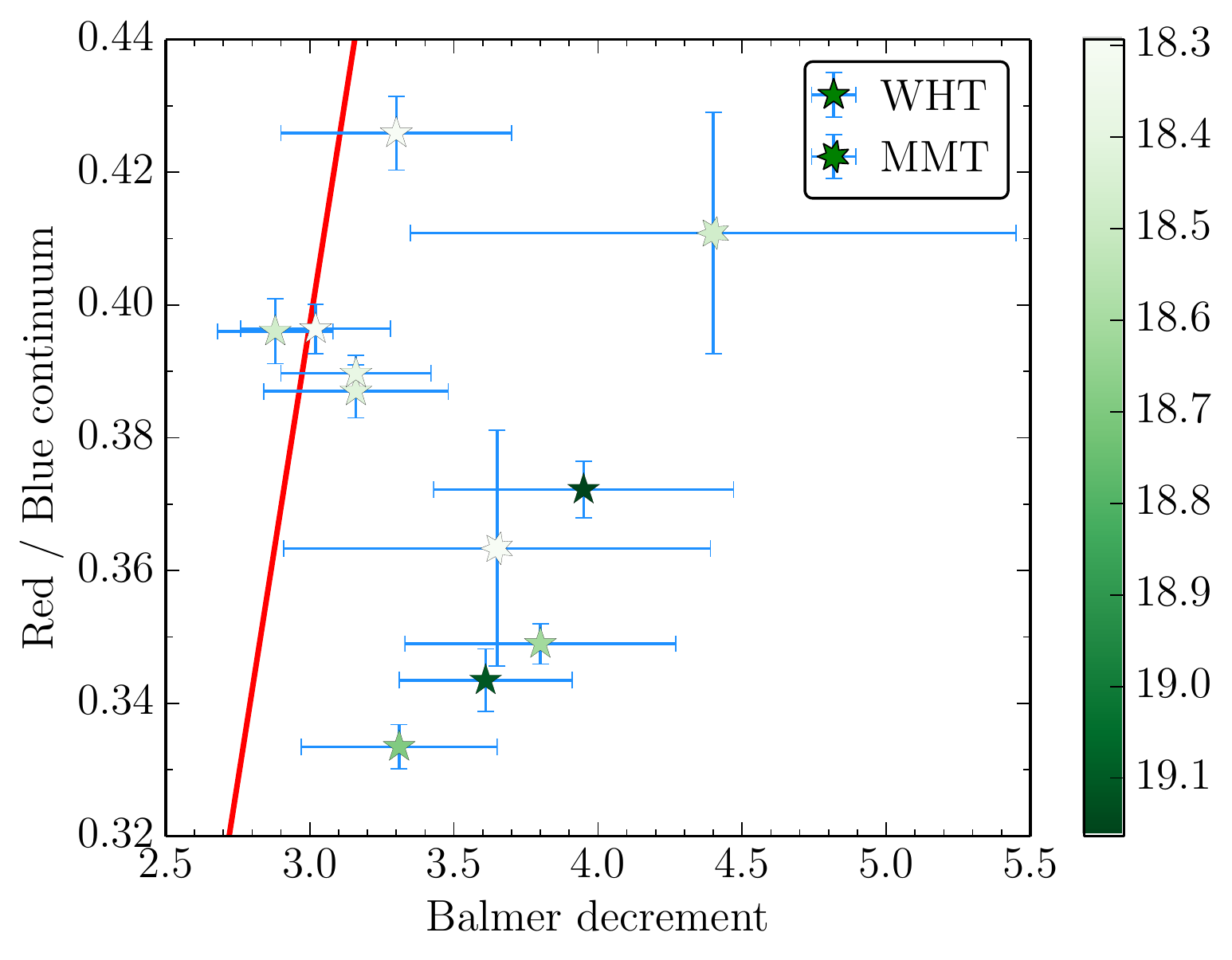} 
\end{center}
\caption{
The Balmer decrement (the ratio of broad H$\upalpha$ to broad H$\upbeta$ fluxes) versus continuum colour (the ratio of $5100\,\si\angstrom$ to $3000\,\si\angstrom$ monochromatic fluxes) as measured in each of the eleven optical spectra taken at the WHT and MMT.  
The colour of the points indicates the equivalent $g$ band magnitude of the spectra calculated in Section~\protect\ref{sec:absflux}: fainter spectra are a darker green.
The red line shows the predicted relation for a \protect\cite{Cardelli89} Galactic reddening curve (assuming the intrinsic Balmer decrement in the case of zero reddening is $2.72$).
}
\label{fig:balmerdec}
\end{figure}

In Figure~\ref{fig:balmerdec} we show how the continuum colour (the ratio of red to blue fluxes) and Balmer decrement have varied together.  
In the simple scenario of a reddening screen of variable column density obscuring both the BLR and accretion disc, there would be a linear relationship between the Balmer decrement and red/blue continuum flux.
We show a reddening vector describing the predicted relationship, again based on the Galactic reddening curve of \cite{Cardelli89} and  positioned so that the Balmer decrement in the case of zero reddening is 2.72 (\citealt{Gaskell17}).
It can be seen that our data do not follow the trend of this reddening vector so reddening alone cannot explain the observed spectral changes.

\subsubsection{Cloud crossing timescale}
If the dimming of the AGN continuum and broad emission line fluxes is due to an obscurer moving across our line of sight, then we can predict the timescale on which such an occultation event would occur.
We estimate the BLR size from \cite{Bentz13} using the equation
\begin{equation}
\log\left(\frac{R_\mathrm{BLR}}{1~\mathrm{light~day}}\right) = K + \alpha \log\left(\frac{\lambda L_{5100\si{\angstrom}}}{10^{44}~\mathrm{erg~s}^{-1}}\right),
\label{eqn:rblr}
\end{equation}
with values $K=1.559$ and $\alpha=0.549$ taken from their `Clean2$+$ExtCor' calibration.
For the range of $\lambda L_{5100\si{\angstrom}}$ observed in our monitoring campaign, the BLR size is $\approx40$--60~light days.

Following \cite{LaMassa15} we calculate the crossing time $t_\mathrm{cross}$ of a cloud occulting the central regions as
\begin{equation}
t_\mathrm{cross} = 0.07 \left(\frac{R_\mathrm{orb}}{1~\mathrm{light~day}}\right)^{\nicefrac{3}{2}} \left(\frac{10^8~\mathrm{M}_{\sun}}{M_\mathrm{BH}}\right)^{\nicefrac{1}{2}} \arcsin\left(\frac{R_\mathrm{src}}{R_\mathrm{orb}}\right)~\mathrm{years}, 
\label{eqn:tcross}
\end{equation}
where $R_\mathrm{orb}$ is the orbital radius of the cloud and $R_\mathrm{src}$ is the radius of the emission source being obscured (here the BLR).
As a conservative estimate (minimising the crossing time), we calculate the crossing time for a cloud at the inner edge of the BLR, i.e.\ $R_\mathrm{src}=R_\mathrm{orb}=R_\mathrm{BLR}\approx50$~light days.
The cloud crossing time at this radius is $\approx27$~years, much longer than the dip-and-rise event we observe in the lightcurve which takes $\approx3$~years in the rest frame.

\subsection{Intrinsic change interpretation}
\subsubsection{Dust reverberation}
\label{sec:dust}
As noted in Section~\ref{sec:wise}, there is a dip in the infrared lightcurves, delayed with respect to the optical dip by around 400~days. 
It can be seen in Figure~\ref{fig:sed} that there is negligible host galaxy emission at the wavelengths of the \textit{WISE} W1 and W2 bands (this is true even in the case of a starburst host galaxy, as the IR emission of starlight-heated dust peaks at longer wavelengths).
The infrared lightcurves may therefore be evidence of AGN-heated dust reverberating with the variable intrinsic AGN continuum.
However, whilst there is a large (factor $\approx3$) change in the optical flux, the change in the near-infrared is much more modest ($\approx30$~per cent).
The dust emission ought to be a good bolometer of the intrinsic AGN luminosity, so we might expect it to show variability of the same amplitude as seen in the optical. 
If we attribute the infrared variability to an echo response to variations in the central source, we must account for this discrepancy.
Here, we assess whether the observed infrared lag and magnitude changes can be plausibly attributed to dust reverberation.

We can calculate the expected dust reverberation radius from our model SED parameters via
\begin{equation}
R_\mathrm{dust,rev} = \sqrt{\frac{L_\mathrm{UV}}{16\uppi\sigma T^4}},
\end{equation}
where $\sigma$ is the Stefan-Boltzmann constant 
and the $T$ is the dust temperature $T=1140$~K (from Table~\ref{tab:sed}).
Since the dust reverberates with the dip in the optical/UV continuum, we take the UV luminosity in the dip to be a factor 2.5 less than we determined at the time of the \textit{XMM-Newton} observation: $L_\mathrm{UV,dip}\approx7.5\times10^{44}$~erg~s$^{-1}$.
We therefore calculate $R_\mathrm{dust,rev}\approx150$~light days.
The observed delay between the minimum of the infrared lightcurve with respect to the optical is $\approx428$~days, equivalent to $\approx335$~days in the rest frame, around a factor two greater than $R_\mathrm{dust,rev}$.

We employ the model tori of \cite{Almeyda17} to simulate how the dust may respond to a variable, driving optical source.
The authors consider the cases of a compact and extended torus, in which the ratio of outer to inner dust cloud radii are 2 and 10, respectively and the inner dust radius in their model is set by dust sublimation.
They consider the effects of differing illumination of the torus dust clouds.
In the case of isotropic illumination, dust sublimation surface is spherical.  
In the anisotropically-illuminated case, more ionising flux is emitted in polar directions than in the equatorial plane; the resultant dust sublimation surface is `bowl-shaped' (see e.g.\ \citealt{Kawaguchi10}) and the dust near the equatorial plane can survive much closer to the central source than in the isotropic case.
The inner dust radius is dependent on the AGN luminosity and the dust sublimation temperature, for which we adopt a value of 1500~K, close to the mean hot dust temperature found by \cite{Landt11}.
For SDSS\,J2232$-$0806, we calculate the AGN luminosity $L_\mathrm{AGN}$ from the bolometric flux of our model SED and assume that was 30 per cent greater in the bright state than observed at the time of the \textit{XMM-Newton} observation.
We therefore determine that $L_\mathrm{AGN}\approx4\times10^{45}$~erg~s$^{-1}$.
For the isotropic case, $R_\mathrm{in}=R_\mathrm{sub}\approx0.7$~pc ($\approx800$ light days) whereas for the anisotropic case $R_\mathrm{in}=R_\mathrm{sub}(\theta=90^\circ)\approx0.25$~pc ($\approx300$ light days).  

\begin{figure}
\begin{center}
\includegraphics[width=\columnwidth,keepaspectratio]{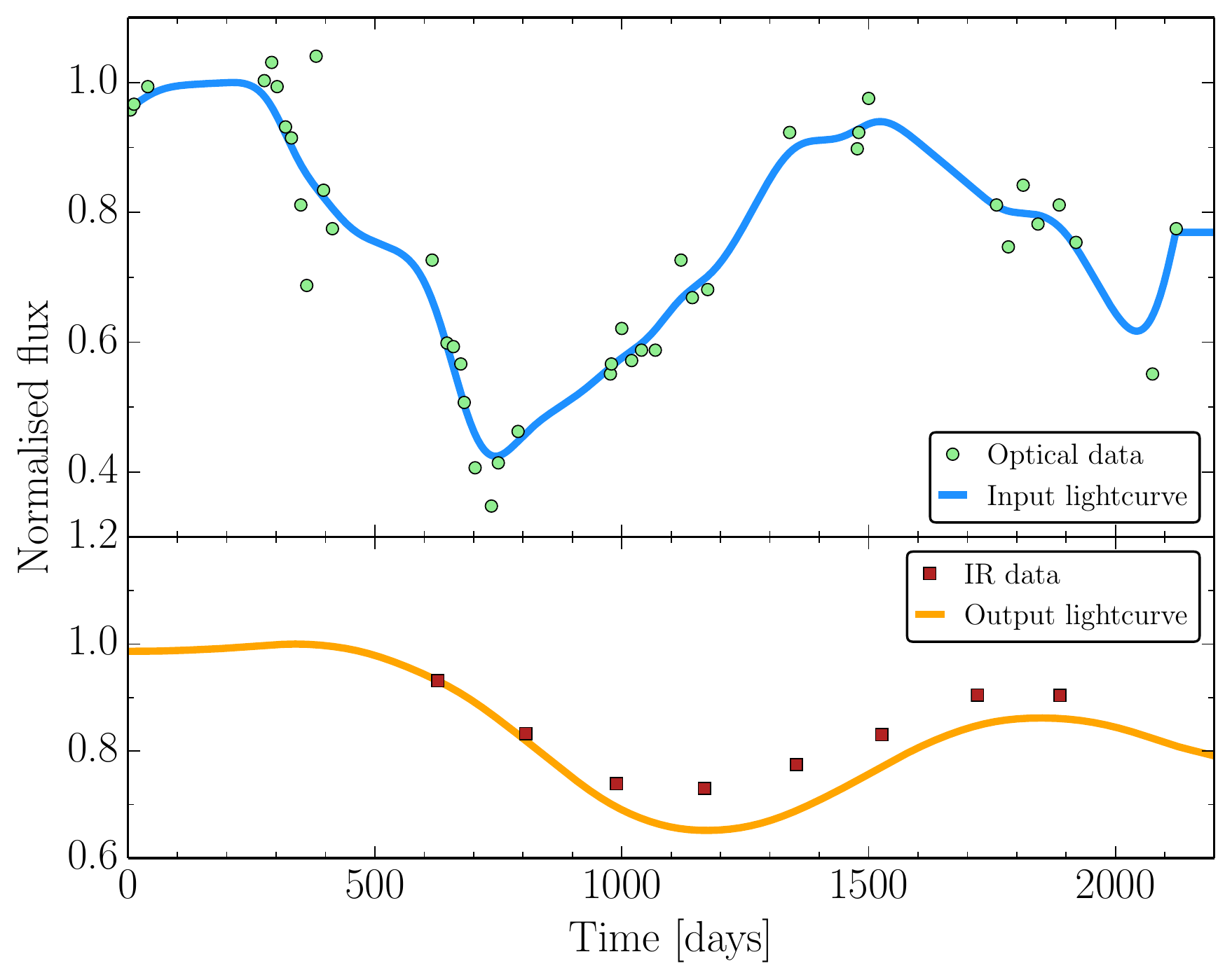} 
\end{center}
\caption{
The simulated dust response to the variable optical source.
\textit{Top:} Our optical data (LT $g$ band photometry) are shown as green circles.
We linearly interpolate between these and smooth the result to create an input optical lightcurve (the blue line).   
\textit{Bottom:} 
We convolve the input lightcurve with an impulse response function to predict the infrared lightcurve (the orange line).
The impulse response function was calculated by \protect\cite{Almeyda17} for an anisotropically-illuminated, radially extended ($R_\mathrm{out}/R_\mathrm{in}=10$) distribution of dust clouds in a torus of angular width $\sigma=45^\circ$ with $R_\mathrm{dust}=250$ light days and a viewed at a polar angle of $\theta_\mathrm{obs}=45^\circ$.
The \textit{WISE} W1 (3.4~$\upmu$m) photometry are shown as red squares and the data is normalised such that the first point falls on the predicted lightcurve.}
\label{fig:torus}
\end{figure}

To construct our driving lightcurve, we interpolate between the LT optical photometry points to create a continuous lightcurve, which we then smooth to remove the short-term, stochastic variability and retain only the shape of the longer-term, systematic, large-amplitude changes.
\cite{Almeyda17} provide their impulse response functions at 3.6~$\upmu$m for a torus viewed at a polar angle of $\theta_\mathrm{obs}=45^\circ$ (see their Figure~8).
We convolve our optical lightcurve with four response functions (for compact/extended, isotropically-/anisotropically-illuminated tori) and compare the simulated dust responses with our \textit{WISE} W1 data.
We find that the response functions for the isotropically-illuminated tori produce much longer lags than is observed.
The lags for the anisotropically-illuminated tori are shorter because the of the closer proximity of the dust to the optical/UV source, and are in much better agreement with our data.
The simulated responses for the compact tori are too deep, an extended distribution is required to smear out the response and reduce its amplitude.
In Figure~\ref{fig:torus} we show the simulated dust response in the case of an extended, anisotropically-illuminated torus.
In this figure we have slightly decreased $R_\mathrm{in}$ to 250~light days from the 300~light days calculated from $L_\mathrm{AGN}$, to better match the observed lightcurve.  

\subsubsection{Accretion disc variability timescales}
\label{sec:inttime}
We now assess the predicted timescales for the transmission of changes through a standard thin accretion disc.
In Section~\ref{sec:optxagnf} we determined the outer radius of the accretion disc to be $R\sim100~R_\mathrm{g}$.
For a disc of this size, the dynamical timescale is
\begin{equation}
t_\mathrm{dyn} \approx \left(\frac{R^3}{GM_\mathrm{BH}}\right)^{\nicefrac{1}{2}} \approx 10~\mathrm{days};
\end{equation}
the thermal timescale is
\begin{equation}
t_\mathrm{therm} \approx \frac{t_\mathrm{dyn}}{\alpha} \approx 3~\mathrm{months},
\end{equation}
where $\alpha\approx0.1$ is the disc viscosity parameter; the viscous timescale is 
\begin{equation}
t_\mathrm{visc} \approx \frac{t_\mathrm{dyn}}{\alpha}\left(\frac{H}{R}\right)^{-2} \approx 1~\mathrm{Gyr}
\end{equation}
where $H/R$ is the ratio of the disc's thickness to its radius. 
As \cite{Noda18} found for Mrk\,1018, we find that for SDSS\,J2232$-$0806 the dynamical and thermal timescales are too short compared with the observed variability timescale and the viscous timescale is far too long.


\section{Discussion}
\label{sec:disc}
\subsection{An extrinsic cause of variability}
The hypothesis of an extrinsic cause of the variability (i.e.\ variable obscuration) is inconsistent with the observations in several important respects:
\begin{itemize}
\item The continuum colour change is inconsistent with reddening since we see approximately equal fractional flux change in the red as in the blue (Figure~\ref{fig:var}).
Even if we perform no subtraction of host galaxy flux at 5100~\AA\, the source still exhibits significantly more variability in the red than would be inferred from the blue, assuming that reddening causes the variability.
We note that the choice of reddening curve makes very little difference at the wavelengths we studied.
\item The Balmer decrements do not change consistently (Figure~\ref{fig:balmerdec}), although this test is less compelling given the substantial uncertainties in the measurements.  
However, since the emission line EWs change, the obscurer cannot be covering both the accretion disc and all of the BLR.
\item We are able to place an upper limit of $7\times10^{19}$~cm$^{-2}$ on the intrinsic column density from the \textit{XMM-Newton} X-ray observation, although a column of $\approx4\times10^{20}$~cm$^{-2}$ would be required to produce the observed 30 per cent drop in the $g$ band flux.
Furthermore, \cite{Maiolino01} reported that the dust reddening of AGN is generally much lower than one would calculate from the gas column density probed by X-rays, assuming a Galactic dust-to-gas ratio and extinction curve, as we do here.
If this were the case for SDSS\,J2232$-$0806, an even greater $N^\mathrm{int}_\mathrm{H}$ would be predicted, increasing the discrepancy with the X-ray observations.
\item The timescale for obscuration is far too long.
We calculate that the crossing time of an obscuring cloud at the inner BLR radius is $\approx27$~years, much longer than the 3~years we observe.
Furthermore, this scenario does not explain how a dust cloud could survive relatively near to the central ionising source.
\item Variable obscuration fails to explain the observed variations in the infrared.  
Since mid-infrared wavelengths are less sensitive to reddening than the optical, a 0.26~mag change at 3.4~$\upmu$m would imply a simultaneous $5.5$~mag change in the $g$ band which is clearly inconsistent with out data.
If the obscurer were exterior to the torus, it would be at an extremely large orbital radius and the crossing time would be even longer than calculated above.
If the obscurer were interior to the torus it would need to be implausibly close to the accretion disc (to explain the lag), and implausibly large (to obscure a sufficient fraction of the AGN flux as seen by the dust).
\end{itemize}

\subsection{An inrinsic cause of variability}
Having ruled out the possibility of an extrinsic change, we consider that the variability is due to an intrinsic change in the luminosity of the accreting matter.
In Section~\ref{sec:dust} we simulated dust responses to a driving optical continuum.
Our intention with this test was not to infer the properties of the torus but to examine the plausibility that the infrared emission reverberates with the optical.
Although we have tested only a few points in the dust response parameter space presented by \cite{Almeyda17}, the simulated IR lightcurve shown in Figure~\ref{fig:torus} captures both the lag and shape of the observed IR variability very well.
It is therefore very plausible that the IR emission exhibits a genuine light echo of the optical variability.

\cite{Sheng17} studied a sample of changing-look quasars that exhibited significant, large-amplitude ($|\Delta \mathrm{W}1|$ or $|\Delta \mathrm{W}2|>0.4$~mag) mid-infrared variability.
Since mid-infrared wavelengths are not strongly affected by dust extinction, the mid-infrared variability would imply much greater changes in the optical than observed if both were due to variable obscuration.
They also found that the timescales for dust cloud obscuration of the torus were far too long whereas the observed lags between infrared and optical were consistent with those expected for hot dust reverberation.
They concluded that in all of the ten objects they investigated that the variability was intrinsic in nature.
We argue that SDSS\,J2232$-$0806 shows the same behaviour.

The RMS spectrum (Figure~\ref{fig:rms}) indicates different variability behaviours of the observed emission lines.
The broad Balmer emission lines appear as strong features in the RMS spectrum, highlighting their significant variability. 
The He\,\textsc{ii}~$\lambda4686$ and He\,\textsc{i}~$\lambda5876$ emission lines are known to respond strongly and rapidly to changes in the continuum, and are also prominent.
However, Mg\,\textsc{ii} almost completely disappears in the RMS spectrum, indicating that it has varied very little over our monitoring campaign.
Both \cite{Zhu17} and \cite{Sun15} have studied the reverberation of Mg\,\textsc{ii} in quasars observed multiple times as part of the SDSS.
\cite{Zhu17} noted that Mg\,\textsc{ii} responds relatively weakly to changes in the 3000~\AA\ continuum.
\cite{Sun15} compared the Mg\,\textsc{ii} and H$\upbeta$ emission line variability and found that Mg\,\textsc{ii} is $\approx1.5$ times less responsive to changes in the continuum than H$\upbeta$.
It is not currently known why this is the case.
It may be that Mg\,\textsc{ii} is emitted over a much larger range of radii than H$\upbeta$ and so its response is more strongly diluted.
Alternatively, differences in the excitation/de-excitation mechanisms or in the optical depths of the two lines are also possible explanations.

Whilst we favour an intrinsic cause of the variability over an extrinsic cause, our calculations in Section~\ref{sec:inttime} show that the predicted timescales for such changes do not match the observations.
It has been known for some time that large-amplitude variability of AGN occurs on timescales much shorter than predicted for thin, viscous accretion discs.  
\cite{Dexter19} address this so-called `quasar viscosity crisis' (\citealt{Lawrence18}) and propose that all AGN accretion discs may be `magnetically elevated' and have a much greater scale height than is typically assumed, dramatically reducing the predicted variability timescales.
Alternative models have recently been developed to explain the extreme variability seen in individual sources.
\cite{Ross18} presented a scenario for the CLQ SDSS\,J110057.70$-$005304.5 in which a dramatic change in magnetic torque at the innermost disc radii resulted in a collapse of the UV continuum and triggered a cooling/heating front propagating through the disc, out to $\sim200~R_\mathrm{g}$.
Taking a different approach, \cite{Noda18} determined that Mrk\,1018 underwent a spectral state transition , similar to those seen in stellar-mass black hole binaries (BHBs).
Whilst scaling up to AGN size-scales by BH mass predicts too long variability timescales in AGN, the authors discuss ways in which scalings between BHBs and AGN may break down.

\subsection{SDSS\,J2232$-$0806 in the context of other hypervaribale AGN}
Both \cite{MacLeod18} and \cite{Rumbaugh18} have recently presented the results of systematic searches of long-term extremely variable quasars  
(EVQs: sources with $|\Delta g|>1$~mag) from archival optical data.
\cite{Rumbaugh18} found that EVQs account for $\approx30$--50~per cent of all quasars and that the EVQs had systematically lower $\nicefrac{L}{L_\mathrm{Edd}}$ than the parent sample of `normal' quasars.
\cite{MacLeod18} presented follow-up spectroscopic observations of a sample of EVQs and were able to confirm that $\approx20$~per cent of these were CLQs.
The authors compared the CLQs with a luminosity- and redshift-matched, lesser-variable control sample and again found that CLQs on average have lower $\nicefrac{L}{L_\mathrm{Edd}}$ than their less-variable counterparts.  
Both studies suggested that EVQs and CLQs represent the extremes of a tail of `normal' quasar variability.
At the far range of this tail, some sources exhibit nearly an order of magnitude change in optical flux over a baseline of $\sim10$~years.
Compared to many of these changing-look AGN, the continuum flux change we observed during our monitoring of SDSS\,J2232$-$0806 is modest.
Its $\log(\nicefrac{L}{L_\mathrm{Edd}})=-1$ is slightly higher than the peaks of the distributions of CLQs and EVQs (which occur at $\log(\nicefrac{L}{L_\mathrm{Edd}})\approx-1.5$, see Figure 6 of \citealt{MacLeod18}) although it is consistent with the range of values for all of the populations shown (CLQs, EVQs, the less-variable control sample and all 105783 of the SDSS DR7 quasars).
Assuming the bolometric flux of the source decreases proportionally to the observed optical, we can estimate that the accretion rate of SDSS\,J2232$-$0806 drops to $\sim$a few per cent of Eddington in the faint state.   

\cite{Elitzur09} proposed a disc wind model of the BLR in which AGN with a very low $L/L_\mathrm{Edd}$ are unable to support a BLR.
After studying a sample of low-luminosity AGN, they determined that the BLR disappears when the AGN luminosity drops below a critical value, $L_\mathrm{AGN}\lesssim5\times10^{39}(M_\mathrm{BH}/10^{7}\mathrm{M}_{\sun})^{\nicefrac{2}{3}}$~erg~s$^{-1}$.
\cite{MacLeod18} found that their CLQs were distributed close to this critical value and likely dropped below it in their faint state, naturally explaining the disappearance of the broad emission lines.
Whilst the broad Balmer emission lines in SDSS\,J2232$-$0806 do weaken in response to a dimming continuum, the source does not satisfy the criterion of a changing-look AGN because these lines have not been observed to disappear.
The source was in a deep minimum in 2000, and it is likely that the host galaxy emission dominates all of the SDSS bands except $u$.
Its UV flux in this epoch was $\approx4$ times fainter than when it was observed by \textit{XMM-Newton}.
Assuming the bolometric flux was also 4 times fainter, its luminosity in 2000 was $\approx8\times10^{44}$~erg~s$^{-1}$.
For the BH mass of SDSS\,J2232$-$0806, the critical luminosity for a BLR in the disc-wind model is $\approx4\times10^{40}$~erg~s$^{-1}$, so the broad lines ought to have been visible even in this deep minimum.
Therefore, we suggest that SDSS\,J2232$-$0806 lies on the sequence of quasar variability, being highly variable whilst its mass accretion rate is too high for it to undergo a changing-look transition.

Whereas optically variable AGN are typically `bluer-when-brighter' (e.g.\ \citealt{Rumbaugh18}, \citealt{Wilhite05}) we do not see strong evidence of that behaviour in our optical monitoring of SDSS\,J2232$-$0806.
Having corrected the longer-wavelength fluxes for host galaxy contamination, we show in Figure~\ref{fig:var} that the fractional variabilities in the blue and red are similar (i.e.\ there is no significant colour change).
In Figure~\ref{fig:rms} we show that the shape of the RMS spectrum is very similar to that of the mean spectrum at the shorter wavelengths less affected by host galaxy contamination.
We note that \cite{Wilhite05} used a sample of higher-redshift quasars than SDSS\,J2232$-$0806 ($z>0.5$) so they probed further into the rest frame UV than we do. 
The authors show that there is a spectral break in the variability of their sample around 2500~\AA\ in the rest frame, with wavelengths shorter than this being more strongly variable.
We may not see evidence of a spectral shape change in SDSS\,J2232$-$0806 because our spectra do not sample below 2500~\AA. 
Furthermore, we see in SDSS\,J2232$-$0806 that changes in the red and blue optical continuum (predominantly emitted from larger/smaller radii, respectively) appear to occur in tandem.  
We do not see a delay in the variations between the red and blue optical emission, indicative of a heating/cooling front propagating through the disc, such as in the model described by \cite{Ross18}.

\subsection{Prospects for future work}
Our observing campaign was fortunate to have recorded a dramatic dimming and brightening event of SDSS\,J2232$-$0806.
There is some evidence that similar events have occurred in its past.  
The source appears to have been in a relatively bright state when observed photographically in 1988 but was in a deep minimum in the SDSS observation of 2000.  
The Catalina lightcurve suggests another dip occurred between 2005--2007 (see Figure~\ref{fig:lightcurve}).
As noted by \cite{MacLeod18}, past hypervariable behaviour is an indicator of future events.
Future monitoring of this source is desirable as we may capture other interesting episodes of variability.

New X-ray and UV observations would be highly beneficial in further investigating the nature of the variability.
Sampling both sides of the peak of the accretion disc emission peak the would enable us to parameterise the changing energetics during a variability episode and determine whether SDSS\,J2232$-$0806 undergoes spectral state changes as seen in e.g.\ Mrk\,1018 (\citealt{Noda18}).
The ability of UV and X-rays to probe the innermost regions would enable us to determine whether some `collapse' of the UV emission occurs as seen in e.g.\ SDSS\,J110057.70$-$005304.5 (\citealt{Ross18}).


\section{Conclusions}
Our recent optical photometric and spectroscopic monitoring campaign on the hypervariable AGN SDSS\,J2232$-$0806 has recorded one dimming and brightening episode with a factor $\approx3$ flux change over four years.
Whilst the observed variability of the source is modest compared to that seen in changing-look AGN, it is extreme compared to the broader AGN population.
We have been able to demonstrate that variable obscuration does not explain the observed spectral changes, nor does it fit the observed timescales for variability in the optical or near-infrared.
An intrinsic change in the AGN luminosity is therefore a likelier explanation, although the observed changes are much more rapid than the theoretical accretion disc viscous timescale.
SDSS\,J2232$-$0806 is one of a growing number of objects which challenge our models of viscous accretion discs.
Whilst we are unable to determine the cause of the intrinsic luminosity change, X-ray and UV monitoring of future episodes should greatly improve our understanding of the processes at work.

\section*{Acknowledgements}
DK acknowledges the receipt of a UK Science and Technology Facilities Council (STFC) studentship (ST/N50404X/1). 
DK and MJW acknowledge support from the STFC grant ST/P000541/1.
Thanks to Michael Fausnaugh for his assistance in the use of \textsc{mapspec}.
Thanks also to Ra'ad Mahmoud, Raj Sathyaprakash, Chris Done, David Rosario and Brad Peterson for useful discussions.

In this research we have made use of the following:
\begin{itemize}
\item data from the William Herschel Telescope, operated on the island of La Palma by the Isaac Newton Group of Telescopes in the Spanish Observatorio del Roque de los Muchachos of the Instituto de Astrof\'{i}sica de Canarias;
\item data from the Liverpool Telescope, operated on the island of La Palma by Liverpool John Moores University in the Spanish Observatorio del Roque de los Muchachos of the Instituto de Astrof\'{i}sica de Canarias with financial support from the STFC;
\item observations obtained at the MMT Observatory, a joint facility of the Smithsonian Institution and the University of Arizona;
\item data products from the \textit{WISE} mission,
which is a joint project of the University of California, Los Angeles,
and the Jet Propulsion Laboratory/California Institute of Technology,
and \textit{NEOWISE}, which is a project of the Jet Propulsion Laboratory/California Institute of Technology. 
\textit{WISE} and \textit{NEOWISE} are both funded by the National Aeronautics and Space Administration (NASA);
\item data from and software developed for \textit{XMM-Newton}, an ESA science mission with instruments and contributions directly funded by ESA Member States and NASA;
\item observations made with the NASA \textit{Galaxy Evolution Explorer}. \textit{GALEX} is operated for NASA by the California Institute of Technology under NASA contract NAS5-98034;
\item data products from the Two Micron All Sky Survey (2MASS), which is a joint project of the University of Massachusetts and the Infrared Processing and Analysis Center/California Institute of Technology, funded by NASA and the National Science Foundation;
\item data from SDSS: funding for the SDSS and SDSS-II has been provided by the Alfred P.\ Sloan Foundation, the Participating Institutions, the National Science Foundation, the U.S.\ Department of Energy, the National Aeronautics and Space Administration, the Japanese Monbukagakusho, the Max Planck Society, and the Higher Education Funding Council for England. The SDSS Web Site is \url{http://www.sdss.org/}; 
\item data from Pan-STARRS-1: the Pan-STARRS-1 Surveys (PS1) have been made possible through contributions of the Institute for Astronomy, the University of Hawaii, the Pan-STARRS Project Office, the Max-Planck Society and its participating institutes, the Max Planck Institute for Astronomy, Heidelberg and the Max Planck Institute for Extraterrestrial Physics, Garching, The Johns Hopkins University, Durham University, the University of Edinburgh, Queen's University Belfast, the Harvard-Smithsonian Center for Astrophysics, the Las Cumbres Observatory Global Telescope Network Incorporated, the National Central University of Taiwan, the Space Telescope Science Institute, the National Aeronautics and Space Administration under Grant No.\ NNX08AR22G issued through the Planetary Science Division of the NASA Science Mission Directorate, the National Science Foundation under Grant No.\ AST-1238877, the University of Maryland, and Eotvos Lorand University (ELTE);
\item data obtained from the SuperCOSMOS Science Archive, prepared and hosted by the Wide Field Astronomy Unit, Institute for Astronomy, University of Edinburgh, which is funded by the UK STFC;
\item the \textsc{SpectRes} spectral resampling tool (\citealt{Carnall17});
\item Doug Welch's Excellent Absorption Law Calculator (\url{http://www.dougwelch.org/Acurve.html});
\item Ned Wright's Cosmology Calculator (\citealt{Wright06}). 
\end{itemize}




\bibliographystyle{mnras}
\bibliography{bigdipper-bib} 

\begin{thebibliography}{}
\makeatletter
\relax
\def\mn@urlcharsother{\let\do\@makeother \do\$\do\&\do\#\do\^\do\_\do\%\do\~}
\def\mn@doi{\begingroup\mn@urlcharsother \@ifnextchar [ {\mn@doi@}
  {\mn@doi@[]}}
\def\mn@doi@[#1]#2{\def\@tempa{#1}\ifx\@tempa\@empty \href
  {http://dx.doi.org/#2} {doi:#2}\else \href {http://dx.doi.org/#2} {#1}\fi
  \endgroup}
\def\mn@eprint#1#2{\mn@eprint@#1:#2::\@nil}
\def\mn@eprint@arXiv#1{\href {http://arxiv.org/abs/#1} {{\tt arXiv:#1}}}
\def\mn@eprint@dblp#1{\href {http://dblp.uni-trier.de/rec/bibtex/#1.xml}
  {dblp:#1}}
\def\mn@eprint@#1:#2:#3:#4\@nil{\def\@tempa {#1}\def\@tempb {#2}\def\@tempc
  {#3}\ifx \@tempc \@empty \let \@tempc \@tempb \let \@tempb \@tempa \fi \ifx
  \@tempb \@empty \def\@tempb {arXiv}\fi \@ifundefined
  {mn@eprint@\@tempb}{\@tempb:\@tempc}{\expandafter \expandafter \csname
  mn@eprint@\@tempb\endcsname \expandafter{\@tempc}}}

\bibitem[\protect\citeauthoryear{{Almeyda}, {Robinson}, {Richmond}, {Vazquez}
  \& {Nikutta}}{{Almeyda} et~al.}{2017}]{Almeyda17}
{Almeyda} T.,  {Robinson} A.,  {Richmond} M.,  {Vazquez} B.,   {Nikutta} R.,
  2017, \mn@doi [\apj] {10.3847/1538-4357/aa7687}, \href
  {http://adsabs.harvard.edu/abs/2017ApJ...843....3A} {843, 3}

\bibitem[\protect\citeauthoryear{{Arnaud}}{{Arnaud}}{1996}]{Arnaud96}
{Arnaud} K.~A.,  1996, in {Jacoby} G.~H.,  {Barnes} J.,  eds,  Astronomical
  Society of the Pacific Conference Series Vol. 101, Astronomical Data Analysis
  Software and Systems V. p.~17

\bibitem[\protect\citeauthoryear{{Bentz} et~al.,}{{Bentz}
  et~al.}{2013}]{Bentz13}
{Bentz} M.~C.,  et~al., 2013, \mn@doi [\apj] {10.1088/0004-637X/767/2/149},
  \href {http://adsabs.harvard.edu/abs/2013ApJ...767..149B} {767, 149}

\bibitem[\protect\citeauthoryear{{Bohlin}, {Savage}  \& {Drake}}{{Bohlin}
  et~al.}{1978}]{Bohlin78}
{Bohlin} R.~C.,  {Savage} B.~D.,   {Drake} J.~F.,  1978, \mn@doi [\apj]
  {10.1086/156357}, \href {http://adsabs.harvard.edu/abs/1978ApJ...224..132B}
  {224, 132}

\bibitem[\protect\citeauthoryear{{Bruce} et~al.,}{{Bruce}
  et~al.}{2017}]{Bruce17}
{Bruce} A.,  et~al., 2017, \mn@doi [\mnras] {10.1093/mnras/stx168}, \href
  {http://adsabs.harvard.edu/abs/2017MNRAS.467.1259B} {467, 1259}

\bibitem[\protect\citeauthoryear{{Bruhweiler} \& {Verner}}{{Bruhweiler} \&
  {Verner}}{2008}]{Bruhweiler08}
{Bruhweiler} F.,  {Verner} E.,  2008, \mn@doi [\apj] {10.1086/525557}, \href
  {http://adsabs.harvard.edu/abs/2008ApJ...675...83B} {675, 83}

\bibitem[\protect\citeauthoryear{{Cannon}}{{Cannon}}{1975}]{Canon75}
{Cannon} R.~D.,  1975, \mn@doi [Proceedings of the Astronomical Society of
  Australia] {10.1017/S1323358000014119}, \href
  {http://adsabs.harvard.edu/abs/1975PASAu...2..323C} {2, 323}

\bibitem[\protect\citeauthoryear{{Cardelli}, {Clayton}  \& {Mathis}}{{Cardelli}
  et~al.}{1989}]{Cardelli89}
{Cardelli} J.~A.,  {Clayton} G.~C.,   {Mathis} J.~S.,  1989, \mn@doi [\apj]
  {10.1086/167900}, \href {http://adsabs.harvard.edu/abs/1989ApJ...345..245C}
  {345, 245}

\bibitem[\protect\citeauthoryear{{Carnall}}{{Carnall}}{2017}]{Carnall17}
{Carnall} A.~C.,  2017, preprint, \href
  {http://adsabs.harvard.edu/abs/2017arXiv170505165C} {} (\mn@eprint {arXiv}
  {1705.05165})

\bibitem[\protect\citeauthoryear{{Chambers} et~al.,}{{Chambers}
  et~al.}{2016}]{Chambers16}
{Chambers} K.~C.,  et~al., 2016, preprint, \href
  {http://adsabs.harvard.edu/abs/2016arXiv161205560C} {} (\mn@eprint {arXiv}
  {1612.05560})

\bibitem[\protect\citeauthoryear{{Collinson} et~al.,}{{Collinson}
  et~al.}{2018}]{Collinson18}
{Collinson} J.~S.,  et~al., 2018, \mn@doi [\mnras] {10.1093/mnras/stx2992},
  \href {http://adsabs.harvard.edu/abs/2018MNRAS.474.3565C} {474, 3565}

\bibitem[\protect\citeauthoryear{{Dexter} \& {Begelman}}{{Dexter} \&
  {Begelman}}{2019}]{Dexter19}
{Dexter} J.,  {Begelman} M.~C.,  2019, \mn@doi [\mnras]
  {10.1093/mnrasl/sly213}, \href
  {http://adsabs.harvard.edu/abs/2019MNRAS.483L..17D} {483, L17}

\bibitem[\protect\citeauthoryear{{Dickey} \& {Lockman}}{{Dickey} \&
  {Lockman}}{1990}]{DL90}
{Dickey} J.~M.,  {Lockman} F.~J.,  1990, \mn@doi [\araa]
  {10.1146/annurev.aa.28.090190.001243}, \href
  {http://adsabs.harvard.edu/abs/1990ARA%26A..28..215D} {28, 215}

\bibitem[\protect\citeauthoryear{{Done}, {Davis}, {Jin}, {Blaes}  \&
  {Ward}}{{Done} et~al.}{2012}]{Done12}
{Done} C.,  {Davis} S.~W.,  {Jin} C.,  {Blaes} O.,   {Ward} M.,  2012, \mn@doi
  [\mnras] {10.1111/j.1365-2966.2011.19779.x}, \href
  {http://adsabs.harvard.edu/abs/2012MNRAS.420.1848D} {420, 1848}

\bibitem[\protect\citeauthoryear{{Elitzur} \& {Ho}}{{Elitzur} \&
  {Ho}}{2009}]{Elitzur09}
{Elitzur} M.,  {Ho} L.~C.,  2009, \mn@doi [\apjl]
  {10.1088/0004-637X/701/2/L91}, \href
  {http://adsabs.harvard.edu/abs/2009ApJ...701L..91E} {701, L91}

\bibitem[\protect\citeauthoryear{{Elvis} et~al.,}{{Elvis}
  et~al.}{2012}]{Elvis12}
{Elvis} M.,  et~al., 2012, \mn@doi [\apj] {10.1088/0004-637X/759/1/6}, \href
  {http://adsabs.harvard.edu/abs/2012ApJ...759....6E} {759, 6}

\bibitem[\protect\citeauthoryear{{Fausnaugh}}{{Fausnaugh}}{2017}]{Fausnaugh17}
{Fausnaugh} M.~M.,  2017, \mn@doi [\pasp] {10.1088/1538-3873/129/972/024007},
  \href {http://adsabs.harvard.edu/abs/2017PASP..129b4007F} {129, 024007}

\bibitem[\protect\citeauthoryear{{Gaskell}}{{Gaskell}}{2017}]{Gaskell17}
{Gaskell} C.~M.,  2017, \mn@doi [\mnras] {10.1093/mnras/stx094}, \href
  {http://adsabs.harvard.edu/abs/2017MNRAS.467..226G} {467, 226}

\bibitem[\protect\citeauthoryear{{Gezari} et~al.,}{{Gezari}
  et~al.}{2017}]{Gezari17}
{Gezari} S.,  et~al., 2017, \mn@doi [\apj] {10.3847/1538-4357/835/2/144}, \href
  {http://adsabs.harvard.edu/abs/2017ApJ...835..144G} {835, 144}

\bibitem[\protect\citeauthoryear{{Goodrich}}{{Goodrich}}{1995}]{Goodrich95}
{Goodrich} R.~W.,  1995, \mn@doi [\apj] {10.1086/175256}, \href
  {http://adsabs.harvard.edu/abs/1995ApJ...440..141G} {440, 141}

\bibitem[\protect\citeauthoryear{{Hambly} et~al.,}{{Hambly}
  et~al.}{2001}]{Hambly01}
{Hambly} N.~C.,  et~al., 2001, \mn@doi [\mnras]
  {10.1111/j.1365-2966.2001.04660.x}, \href
  {http://adsabs.harvard.edu/abs/2001MNRAS.326.1279H} {326, 1279}

\bibitem[\protect\citeauthoryear{{Jin}, {Ward}  \& {Done}}{{Jin}
  et~al.}{2012}]{JinWDG12_II}
{Jin} C.,  {Ward} M.,   {Done} C.,  2012, \mn@doi [\mnras]
  {10.1111/j.1365-2966.2012.20847.x}, \href
  {http://adsabs.harvard.edu/abs/2012MNRAS.422.3268J} {422, 3268}

\bibitem[\protect\citeauthoryear{{Katebi} et~al.,}{{Katebi}
  et~al.}{2018}]{Katebi18}
{Katebi} R.,  et~al., 2018, preprint, \href
  {http://adsabs.harvard.edu/abs/2018arXiv181103694K} {} (\mn@eprint {arXiv}
  {1811.03694})

\bibitem[\protect\citeauthoryear{{Kawaguchi} \& {Mori}}{{Kawaguchi} \&
  {Mori}}{2010}]{Kawaguchi10}
{Kawaguchi} T.,  {Mori} M.,  2010, \mn@doi [\apjl]
  {10.1088/2041-8205/724/2/L183}, \href
  {http://adsabs.harvard.edu/abs/2010ApJ...724L.183K} {724, L183}

\bibitem[\protect\citeauthoryear{{Kubota} \& {Done}}{{Kubota} \&
  {Done}}{2018}]{Kubota18}
{Kubota} A.,  {Done} C.,  2018, \mn@doi [\mnras] {10.1093/mnras/sty1890}, \href
  {http://adsabs.harvard.edu/abs/2018MNRAS.480.1247K} {480, 1247}

\bibitem[\protect\citeauthoryear{{LaMassa} et~al.,}{{LaMassa}
  et~al.}{2015}]{LaMassa15}
{LaMassa} S.~M.,  et~al., 2015, \mn@doi [\apj] {10.1088/0004-637X/800/2/144},
  \href {http://adsabs.harvard.edu/abs/2015ApJ...800..144L} {800, 144}

\bibitem[\protect\citeauthoryear{{Landt}, {Elvis}, {Ward}, {Bentz}, {Korista}
  \& {Karovska}}{{Landt} et~al.}{2011}]{Landt11}
{Landt} H.,  {Elvis} M.,  {Ward} M.~J.,  {Bentz} M.~C.,  {Korista} K.~T.,
  {Karovska} M.,  2011, \mn@doi [\mnras] {10.1111/j.1365-2966.2011.18383.x},
  \href {http://adsabs.harvard.edu/abs/2011MNRAS.414..218L} {414, 218}

\bibitem[\protect\citeauthoryear{{Lawrence}}{{Lawrence}}{2018}]{Lawrence18}
{Lawrence} A.,  2018, \mn@doi [Nature Astronomy] {10.1038/s41550-017-0372-1},
  \href {http://adsabs.harvard.edu/abs/2018NatAs...2..102L} {2, 102}

\bibitem[\protect\citeauthoryear{{Lawrence} et~al.,}{{Lawrence}
  et~al.}{2016}]{Lawrence16}
{Lawrence} A.,  et~al., 2016, \mn@doi [\mnras] {10.1093/mnras/stw1963}, \href
  {http://adsabs.harvard.edu/abs/2016MNRAS.463..296L} {463, 296}

\bibitem[\protect\citeauthoryear{{Lu}, {Zhao}, {Bai}  \& {Fan}}{{Lu}
  et~al.}{2019}]{Lu19}
{Lu} K.-X.,  {Zhao} Y.,  {Bai} J.-M.,   {Fan} X.-L.,  2019, \mn@doi [\mnras]
  {10.1093/mnras/sty3229}, \href
  {http://adsabs.harvard.edu/abs/2019MNRAS.483.1722L} {483, 1722}

\bibitem[\protect\citeauthoryear{{MacLeod} et~al.,}{{MacLeod}
  et~al.}{2012}]{MacLeod12}
{MacLeod} C.~L.,  et~al., 2012, \mn@doi [\apj] {10.1088/0004-637X/753/2/106},
  \href {http://adsabs.harvard.edu/abs/2012ApJ...753..106M} {753, 106}

\bibitem[\protect\citeauthoryear{{MacLeod} et~al.,}{{MacLeod}
  et~al.}{2016}]{MacLeod16}
{MacLeod} C.~L.,  et~al., 2016, \mn@doi [\mnras] {10.1093/mnras/stv2997}, \href
  {http://adsabs.harvard.edu/abs/2016MNRAS.457..389M} {457, 389}

\bibitem[\protect\citeauthoryear{{MacLeod} et~al.,}{{MacLeod}
  et~al.}{2018}]{MacLeod18}
{MacLeod} C.~L.,  et~al., 2018, preprint, \href
  {http://adsabs.harvard.edu/abs/2018arXiv181000087M} {} (\mn@eprint {arXiv}
  {1810.00087})

\bibitem[\protect\citeauthoryear{{Magnier} et~al.,}{{Magnier}
  et~al.}{2013}]{Magnier13}
{Magnier} E.~A.,  et~al., 2013, \mn@doi [\apjs] {10.1088/0067-0049/205/2/20},
  \href {http://adsabs.harvard.edu/abs/2013ApJS..205...20M} {205, 20}

\bibitem[\protect\citeauthoryear{{Maiolino}, {Marconi}, {Salvati}, {Risaliti},
  {Severgnini}, {Oliva}, {La Franca}  \& {Vanzi}}{{Maiolino}
  et~al.}{2001}]{Maiolino01}
{Maiolino} R.,  {Marconi} A.,  {Salvati} M.,  {Risaliti} G.,  {Severgnini} P.,
  {Oliva} E.,  {La Franca} F.,   {Vanzi} L.,  2001, \mn@doi [\aap]
  {10.1051/0004-6361:20000177}, \href
  {http://adsabs.harvard.edu/abs/2001A%26A...365...28M} {365, 28}

\bibitem[\protect\citeauthoryear{{Martin} et~al.,}{{Martin}
  et~al.}{2005}]{Galex05}
{Martin} D.~C.,  et~al., 2005, \mn@doi [\apjl] {10.1086/426387}, \href
  {http://adsabs.harvard.edu/abs/2005ApJ...619L...1M} {619, L1}

\bibitem[\protect\citeauthoryear{{McHardy} et~al.,}{{McHardy}
  et~al.}{2018}]{McHardy18}
{McHardy} I.~M.,  et~al., 2018, \mn@doi [\mnras] {10.1093/mnras/sty1983}, \href
  {http://adsabs.harvard.edu/abs/2018MNRAS.480.2881M} {480, 2881}

\bibitem[\protect\citeauthoryear{{Mej{\'{\i}}a-Restrepo}, {Trakhtenbrot},
  {Lira}, {Netzer}  \& {Capellupo}}{{Mej{\'{\i}}a-Restrepo}
  et~al.}{2016}]{Mejia16}
{Mej{\'{\i}}a-Restrepo} J.~E.,  {Trakhtenbrot} B.,  {Lira} P.,  {Netzer} H.,
  {Capellupo} D.~M.,  2016, \mn@doi [\mnras] {10.1093/mnras/stw568}, \href
  {http://adsabs.harvard.edu/abs/2016MNRAS.460..187M} {460, 187}

\bibitem[\protect\citeauthoryear{{Noda} \& {Done}}{{Noda} \&
  {Done}}{2018}]{Noda18}
{Noda} H.,  {Done} C.,  2018, \mn@doi [\mnras] {10.1093/mnras/sty2032}, \href
  {http://adsabs.harvard.edu/abs/2018MNRAS.480.3898N} {480, 3898}

\bibitem[\protect\citeauthoryear{{Osterbrock} \& {Ferland}}{{Osterbrock} \&
  {Ferland}}{2006}]{O&F06}
{Osterbrock} D.~E.,  {Ferland} G.~J.,  2006, {Astrophysics of gaseous nebulae
  and active galactic nuclei}, 2nd edn.
University Science Books, Mill Valley, CA

\bibitem[\protect\citeauthoryear{{Parker} et~al.,}{{Parker}
  et~al.}{2016}]{Parker16}
{Parker} M.~L.,  et~al., 2016, \mn@doi [\mnras] {10.1093/mnras/stw1449}, \href
  {http://adsabs.harvard.edu/abs/2016MNRAS.461.1927P} {461, 1927}

\bibitem[\protect\citeauthoryear{{Pei} et~al.,}{{Pei} et~al.}{2017}]{Pei17}
{Pei} L.,  et~al., 2017, \mn@doi [\apj] {10.3847/1538-4357/aa5eb1}, \href
  {http://adsabs.harvard.edu/abs/2017ApJ...837..131P} {837, 131}

\bibitem[\protect\citeauthoryear{{Peterson} et~al.,}{{Peterson}
  et~al.}{2004}]{Peterson04}
{Peterson} B.~M.,  et~al., 2004, \mn@doi [\apj] {10.1086/423269}, \href
  {http://adsabs.harvard.edu/abs/2004ApJ...613..682P} {613, 682}

\bibitem[\protect\citeauthoryear{{Polletta} et~al.,}{{Polletta}
  et~al.}{2007}]{Polletta07}
{Polletta} M.,  et~al., 2007, \mn@doi [\apj] {10.1086/518113}, \href
  {http://adsabs.harvard.edu/abs/2007ApJ...663...81P} {663, 81}

\bibitem[\protect\citeauthoryear{{Ross} et~al.,}{{Ross} et~al.}{2018}]{Ross18}
{Ross} N.~P.,  et~al., 2018, \mn@doi [\mnras] {10.1093/mnras/sty2002}, \href
  {http://adsabs.harvard.edu/abs/2018MNRAS.480.4468R} {480, 4468}

\bibitem[\protect\citeauthoryear{{Ruan} et~al.,}{{Ruan} et~al.}{2016}]{Ruan16}
{Ruan} J.~J.,  et~al., 2016, \mn@doi [\apj] {10.3847/0004-637X/826/2/188},
  \href {http://adsabs.harvard.edu/abs/2016ApJ...826..188R} {826, 188}

\bibitem[\protect\citeauthoryear{{Rumbaugh} et~al.,}{{Rumbaugh}
  et~al.}{2018}]{Rumbaugh18}
{Rumbaugh} N.,  et~al., 2018, \mn@doi [\apj] {10.3847/1538-4357/aaa9b6}, \href
  {http://adsabs.harvard.edu/abs/2018ApJ...854..160R} {854, 160}

\bibitem[\protect\citeauthoryear{{Runnoe} et~al.,}{{Runnoe}
  et~al.}{2016}]{Runnoe16}
{Runnoe} J.~C.,  et~al., 2016, \mn@doi [\mnras] {10.1093/mnras/stv2385}, \href
  {http://adsabs.harvard.edu/abs/2016MNRAS.455.1691R} {455, 1691}

\bibitem[\protect\citeauthoryear{{Schmidt}, {Rix}, {Shields}, {Knecht}, {Hogg},
  {Maoz}  \& {Bovy}}{{Schmidt} et~al.}{2012}]{Schmidt12}
{Schmidt} K.~B.,  {Rix} H.-W.,  {Shields} J.~C.,  {Knecht} M.,  {Hogg} D.~W.,
  {Maoz} D.,   {Bovy} J.,  2012, \mn@doi [\apj] {10.1088/0004-637X/744/2/147},
  \href {http://adsabs.harvard.edu/abs/2012ApJ...744..147S} {744, 147}

\bibitem[\protect\citeauthoryear{{Sheng}, {Wang}, {Jiang}, {Yang}, {Yan}, {Dou}
   \& {Peng}}{{Sheng} et~al.}{2017}]{Sheng17}
{Sheng} Z.,  {Wang} T.,  {Jiang} N.,  {Yang} C.,  {Yan} L.,  {Dou} L.,   {Peng}
  B.,  2017, \mn@doi [\apjl] {10.3847/2041-8213/aa85de}, \href
  {http://adsabs.harvard.edu/abs/2017ApJ...846L...7S} {846, L7}

\bibitem[\protect\citeauthoryear{{Skrutskie} et~al.,}{{Skrutskie}
  et~al.}{2006}]{Skrutskie06}
{Skrutskie} M.~F.,  et~al., 2006, \mn@doi [\aj] {10.1086/498708}, \href
  {http://adsabs.harvard.edu/abs/2006AJ....131.1163S} {131, 1163}

\bibitem[\protect\citeauthoryear{{Stern} et~al.,}{{Stern}
  et~al.}{2018}]{Stern18}
{Stern} D.,  et~al., 2018, \mn@doi [\apj] {10.3847/1538-4357/aac726}, \href
  {http://adsabs.harvard.edu/abs/2018ApJ...864...27S} {864, 27}

\bibitem[\protect\citeauthoryear{{Sun} et~al.,}{{Sun} et~al.}{2015}]{Sun15}
{Sun} M.,  et~al., 2015, \mn@doi [\apj] {10.1088/0004-637X/811/1/42}, \href
  {http://adsabs.harvard.edu/abs/2015ApJ...811...42S} {811, 42}

\bibitem[\protect\citeauthoryear{{Vanden Berk} et~al.,}{{Vanden Berk}
  et~al.}{2001}]{VandenBerk01}
{Vanden Berk} D.~E.,  et~al., 2001, \mn@doi [\aj] {10.1086/321167}, \href
  {http://adsabs.harvard.edu/abs/2001AJ....122..549V} {122, 549}

\bibitem[\protect\citeauthoryear{{Wang}, {Xu}  \& {Wei}}{{Wang}
  et~al.}{2018}]{Wang18}
{Wang} J.,  {Xu} D.~W.,   {Wei} J.~Y.,  2018, \mn@doi [\apj]
  {10.3847/1538-4357/aab88b}, \href
  {http://adsabs.harvard.edu/abs/2018ApJ...858...49W} {858, 49}

\bibitem[\protect\citeauthoryear{{Wilhite}, {Vanden Berk}, {Kron}, {Schneider},
  {Pereyra}, {Brunner}, {Richards}  \& {Brinkmann}}{{Wilhite}
  et~al.}{2005}]{Wilhite05}
{Wilhite} B.~C.,  {Vanden Berk} D.~E.,  {Kron} R.~G.,  {Schneider} D.~P.,
  {Pereyra} N.,  {Brunner} R.~J.,  {Richards} G.~T.,   {Brinkmann} J.~V.,
  2005, \mn@doi [\apj] {10.1086/430821}, \href
  {http://adsabs.harvard.edu/abs/2005ApJ...633..638W} {633, 638}

\bibitem[\protect\citeauthoryear{{Wright}}{{Wright}}{2006}]{Wright06}
{Wright} E.~L.,  2006, \mn@doi [\pasp] {10.1086/510102}, \href
  {http://adsabs.harvard.edu/abs/2006PASP..118.1711W} {118, 1711}

\bibitem[\protect\citeauthoryear{{Wright} et~al.,}{{Wright}
  et~al.}{2010}]{WISE10}
{Wright} E.~L.,  et~al., 2010, \mn@doi [\aj] {10.1088/0004-6256/140/6/1868},
  \href {http://adsabs.harvard.edu/abs/2010AJ....140.1868W} {140, 1868}

\bibitem[\protect\citeauthoryear{{Yang} et~al.,}{{Yang} et~al.}{2018}]{Yang18}
{Yang} Q.,  et~al., 2018, \mn@doi [\apj] {10.3847/1538-4357/aaca3a}, \href
  {http://adsabs.harvard.edu/abs/2018ApJ...862..109Y} {862, 109}

\bibitem[\protect\citeauthoryear{{Zhu}, {Sun}  \& {Wang}}{{Zhu}
  et~al.}{2017}]{Zhu17}
{Zhu} D.,  {Sun} M.,   {Wang} T.,  2017, \mn@doi [\apj]
  {10.3847/1538-4357/aa76e7}, \href
  {http://adsabs.harvard.edu/abs/2017ApJ...843...30Z} {843, 30}

\bibitem[\protect\citeauthoryear{{Zuo}, {Wu}, {Liu}  \& {Jiao}}{{Zuo}
  et~al.}{2012}]{Zuo12}
{Zuo} W.,  {Wu} X.-B.,  {Liu} Y.-Q.,   {Jiao} C.-L.,  2012, \mn@doi [\apj]
  {10.1088/0004-637X/758/2/104}, \href
  {http://adsabs.harvard.edu/abs/2012ApJ...758..104Z} {758, 104}

\bibitem[\protect\citeauthoryear{{van Groningen} \& {Wanders}}{{van Groningen}
  \& {Wanders}}{1992}]{vanGroningen92}
{van Groningen} E.,  {Wanders} I.,  1992, \mn@doi [\pasp] {10.1086/133039},
  \href {http://adsabs.harvard.edu/abs/1992PASP..104..700V} {104, 700}

\makeatother
\end{thebibliography}



\appendix

\section{Optical spectral fitting results}

\begin{table*}
	\centering
	\caption{Balmer, [O\,\textsc{iii}] and [N\,\textsc{ii}] emission line measurements}
	\label{tab:WHT1}
	\begin{tabular}{lccccccccc}
	\hline
	 & & & & & \multicolumn{5}{c}{H$\upalpha$}   \\
	 & & & & & \multicolumn{5}{c}{$\overbrace{\rule{8.5cm}{0pt}}$}  \\
	Date & Scale & $\Delta v_\mathrm{vb}$ 	& W$_\mathrm{b}$ & W$_\mathrm{n}$ & $f_\mathrm{vb}~\times10^{-15}$ & $f_\mathrm{b}~\times10^{-14}$ & $f_\mathrm{n}~\times10^{-15}$ &  $f_\mathrm{tot}~\times10^{-14}$ & EW$_\mathrm{vb+b}$  \Tstrut\Bstrut \\
	\hline
	\Tstrut 
	2013-06-10 & 1.446 & $+800\pm700$       & $4460\pm50$  & $480\pm10$  & $3\pm2$     & $5.05\pm0.08$ & $1.5\pm0.1$   & $5.5\pm0.2$ & $570\pm20$  		  \\
	2013-08-07 & 1.584 & $+1900\pm700$      & $4530\pm60$  & $550\pm10$  & $3\pm1$     & $5.20\pm0.08$ & $1.8\pm0.1$   & $5.7\pm0.2$ & $750\pm30$  	  \\
	2013-09-09 & 1.046 & $+1300\pm800$      & $4280\pm50$  & $510\pm10$  & $9\pm2$     & $5.4\pm0.1$   & $1.81\pm0.08$ & $6.5\pm0.2$ & $850\pm30$  	 \\
	2014-07-23 & 1.173 & $+800\pm700$       & $4200\pm60$  & $480\pm10$  & $6\pm2$     & $4.97\pm0.09$ & $1.59\pm0.09$ & $5.7\pm0.2$ & $1130\pm50$  	  \\
	2014-12-16 & 1.076 & $+2000\pm500$      & $4300\pm50$  & $570\pm20$  & $5\pm4$     & $4.2\pm0.1$   & $1.7\pm0.1$   & $4.8\pm0.4$ & $1070\pm90$  	  \\
	2016-07-09 & 1.344 & $\leqslant700$     & $4600\pm100$ & $490\pm10$  & $3\pm2$     & $4.8\pm0.1$   & $1.47\pm0.06$ & $5.3\pm0.2$ & $600\pm30$  		 \\
	2016-10-22 & 0.706 & $+1800\pm500$      & $4510\pm40$  & $540\pm10$  & $3\pm1$     & $4.86\pm0.05$ & $1.48\pm0.05$ & $5.3\pm0.1$ & $610\pm10$   	  \\ 
	2017-07-27 & 1.022 & $+1800\pm8000$     & $4500\pm100$ & $470\pm10$  & $4\pm3$     & $5.2\pm0.1$   & $1.43\pm0.07$ & $5.7\pm0.3$ & $970\pm60$  	  \\
	2017-10-11 & 1.152 & $+1900$       		& $3670\pm90$  & $700\pm200$ & $1.2\pm0.3$ & $5\pm1$       & $1.7\pm0.4$   & $7\pm1$     & $1200\pm300$ 	      \\ 
	2017-11-24 & 1.316 & $+1900$            & $4000\pm500$ & $500\pm100$ & $2.1\pm0.7$ & $6\pm1$       & $1.1\pm0.4$   & $8\pm1$     & $1100\pm200$ 	     \\
	2018-07-05 & 1.093 & $+1800\pm800$      & $4560\pm80$  & $490\pm30$  & $4\pm3$     & $4.5\pm0.1$   & $1.5\pm0.1$   & $5.1\pm0.3$ & $660\pm40$  		 \Tstrut \\
	\hline 
	\end{tabular}
\end{table*}

\begin{table*}
	\centering
	\caption{Balmer, [O\,\textsc{iii}] and [N\,\textsc{ii}] emission line measurements (continued)}
	\label{tab:WHT2}
	\begin{tabular}{lccccccccc}
	\hline
	 & \multicolumn{1}{c}{$[\mathrm{N}\textsc{ii}]~\lambda6583$} & \multicolumn{5}{c}{H$\upbeta$} & & \multicolumn{2}{c}{$[\mathrm{O}\textsc{iii}]~\lambda5007$} \\
	 & \multicolumn{1}{c}{$\overbrace{\rule{1.2cm}{0pt}}$} &  \multicolumn{5}{c}{$\overbrace{\rule{8cm}{0pt}}$} 
	 &    
	 & \multicolumn{2}{c}{$\overbrace{\rule{2.5cm}{0pt}}$} \\
	Date & $f~\times10^{-16}$ & $f_\mathrm{vb}~\times10^{-15}$ & $f_\mathrm{b}~\times10^{-14}$ & $f_\mathrm{n}~\times10^{-16}$ & $f_\mathrm{tot}~\times10^{-14}$ & EW$_\mathrm{vb+b}$ & BD$_\mathrm{vb+b}$ & $f~\times10^{-15}$ & EW  
	\Tstrut\Bstrut \\
	\hline
	\Tstrut 
	2013-06-10 & $5.7\pm0.1$ 	& $3\pm1$ & $1.46\pm0.09$ & $2.3\pm0.1$   & $1.8\pm0.1$   	& $112\pm8$   	& $3.0\pm0.3$   	& $3.2\pm0.5$ & $21\pm3$ \\
	2013-08-07 & $6.5\pm0.2$ 	& $3\pm1$ & $1.58\pm0.04$ & $2.7\pm0.2$   & $1.9\pm0.1$   	& $150\pm10$  	& $2.9\pm0.2$ 		& $3.5\pm0.1$ & $28\pm8$ \\
	2013-09-09 & $6.0\pm0.1$ 	& $3\pm2$ & $1.32\pm0.05$ & $2.7\pm0.1$   & $1.7\pm0.2$ 	& $170\pm20$ 	& $3.8\pm0.5$   	& $3.7\pm0.2$ & $40\pm20$ \\
	2014-07-23 & $5.7\pm0.1$ 	& $3\pm1$ & $1.21\pm0.05$ & $2.4\pm0.1$   & $1.6\pm0.1$ 	& $250\pm20$ 	& $3.6\pm0.3$   	& $3.4\pm0.6$ & $56\pm9$ \\
	2014-12-16 & $6.6\pm0.2$ 	& $3\pm1$ & $0.83\pm0.03$ & $2.6\pm0.2$   & $1.2\pm0.1$   	& $190\pm20$ 	& $4.0\pm0.5$   	& $3.3\pm0.8$ & $50\pm10$ \\
	2016-07-09 & $5.7\pm0.2$ 	& $3\pm1$ & $1.32\pm0.09$ & $2.2\pm0.1$   & $1.6\pm0.2$   	& $120\pm10$   	& $3.2\pm0.3$   	& $3.1\pm0.4$ & $24\pm3$ \\
	2016-10-22 & $6.4\pm0.2$ 	& $3\pm1$ & $1.34\pm0.06$ & $2.20\pm0.08$ & $1.7\pm0.1$   	& $110\pm10$   	& $3.2\pm0.3$ 		& $3.3\pm0.2$ & $23\pm2$ \\ 
	2017-07-27 & $5.4\pm0.1$ 	& $3\pm1$ & $1.41\pm0.08$ & $2.1\pm0.1$   & $1.7\pm0.2$ 	& $180\pm20$ 	& $3.3\pm0.3$   	& $3.2\pm0.5$ & $35\pm5$ \\
	2017-10-11 & $8\pm2$ 		& $5\pm2$ & $1.3\pm0.1$   & $2.5\pm0.6$   & $1.8\pm0.2$     & $150\pm20$ 	& $3.7\pm0.7$   	& $2.5\pm0.6$ & $22\pm6$ \\ 
	2017-11-24 & $6\pm2$ 		& $5\pm2$ & $1.3\pm0.2$   & $1.6\pm0.6$   & $1.7\pm0.3$     & $160\pm20$ 	& $4.4\pm1.0$   	& $1.8\pm0.5$ & $16\pm4$ \\
	2018-07-05 & $5.7\pm0.4$	& $3\pm1$ & $1.19\pm0.09$ & $2.2\pm0.2$   & $1.5\pm0.2$   	& $120\pm10$   	& $3.3\pm0.4$   	& $3.4\pm0.8$ & $29\pm6$ \Tstrut \\
	\hline 
	\end{tabular}
	\parbox[]{16.5cm}{`Scale' is the flux scaling factor applied to each spectrum, including both internal and absolute scalings (see Section~\ref{sec:absflux} in the text).  Subscripts `vb', `b' and `n' refer to the very broad, broad and narrow emission line components, respectively and `tot' is the total.  $\Delta v_\mathrm{vb}$ is the velocity offset (in km~s$^{-1}$) of the very broad emission line components relative to the narrower components; positive values indicate a redward offset.  Fluxes $f$ in erg~s$^{-1}$~cm$^{-2}$; widths `W' are FWHM in km~s$^{-1}$ and equivalent widths `EW' are in \AA. `BD' is the Balmer decrement $\nicefrac{\mathrm{H}\upalpha}{\mathrm{H}\upbeta}$.}	
\end{table*}

\begin{table*}
	\centering
	\caption{Mg\,\textsc{ii} emission line measurements}
	\label{tab:WHT-MgII}
	\begin{tabular}{lccccccc}
	\hline
	Date 		& W$_\mathrm{vb}$ & $f_\mathrm{vb}~\times10^{-14}$ & W$_\mathrm{b}$ & $f_\mathrm{b}~\times10^{-14}$ & W$_\mathrm{tot}$ & $f_\mathrm{tot}~\times10^{-14}$ & EW$_\mathrm{tot}$ \\
	\hline
	2013-06-10 	& $9000\pm800$   & $2.2\pm0.5$ & $3300\pm300$ & $1.5\pm0.2$ & $4200\pm400$ & $3.7\pm0.5$ & $51\pm8$   \\
	2013-08-07 	& $9000\pm1000$  & $2.0\pm0.6$ & $3100\pm500$ & $1.5\pm0.3$ & $3900\pm500$ & $3.5\pm0.7$ & $60\pm10$  \\	
	2013-09-09 	& $10000\pm600$  & $2.4\pm0.3$ & $3400\pm100$ & $1.9\pm0.1$ & $4100\pm200$ & $4.3\pm0.3$ & $80\pm10$  \\
	2014-07-23 	& $9000\pm700$   & $1.5\pm0.3$ & $3600\pm200$ & $2.0\pm0.2$ & $4000\pm200$ & $3.5\pm0.4$ & $100\pm10$ \\
	2014-12-16 	& $8400\pm600$   & $2.1\pm0.3$ & $3100\pm200$ & $1.4\pm0.1$ & $4000\pm300$ & $3.5\pm0.3$ & $110\pm10$ \\
	2016-07-09 	& $9000\pm1000$  & $1.9\pm0.5$ & $3300\pm400$ & $1.3\pm0.2$ & $4200\pm500$ & $3.2\pm0.6$ & $50\pm10$  \\
	2016-10-22 	& $11000\pm500$  & $2.4\pm0.3$ & $3400\pm200$ & $1.7\pm0.1$ & $4300\pm200$ & $4.0\pm0.3$ & $59\pm7$   \\
	2017-07-27 	& $9900\pm700$   & $2.0\pm0.3$ & $3600\pm200$ & $2.0\pm0.1$ & $4200\pm200$ & $4.0\pm0.3$ & $73\pm9$   \\
	2017-10-11 	& $14000\pm1300$ & $3.2\pm0.5$ & $3500\pm200$ & $2.3\pm0.2$ & $4200\pm300$ & $5.5\pm0.5$ & $100\pm10$ \\
	2017-11-24 	& $11000\pm1700$ & $2.9\pm0.7$ & $3300\pm300$ & $1.8\pm0.3$ & $4200\pm500$ & $4.7\pm0.7$ & $110\pm20$ \\
	2018-07-05  & $9000\pm1000$  & $2.0\pm0.6$ & $3300\pm400$ & $1.5\pm0.3$ & $4200\pm500$ & $3.5\pm0.7$ & $60\pm10$  \\
	\hline
	\end{tabular}
	\parbox[]{14cm}{Subscripts `vb' and `b' refer to the very broad and broad emission line components, respectively and `tot' is the total.  
	Fluxes $f$ in erg~s$^{-1}$~cm$^{-2}$; widths `W' are FWHM in km~s$^{-1}$ and equivalent widths `EW' are in \AA.}		
\end{table*}


\bsp	
\label{lastpage}
\end{document}